\documentclass[sigconf]{acmart}
\usepackage{graphicx}
\usepackage{amsmath}
\usepackage{amsthm}
\usepackage{booktabs}
\usepackage{algorithm}
\usepackage{algorithmic}
\usepackage{multirow}
\usepackage{subfigure}

\usepackage{multirow} 
\usepackage{booktabs}
\usepackage{ulem}

\AtBeginDocument{%
  }

\copyrightyear{2025}
\acmYear{2025}
\setcopyright{cc}
\setcctype{by}
\acmConference[KDD '25]{Proceedings of the 31st ACM SIGKDD Conference on
Knowledge Discovery and Data Mining V.1}{August 3--7, 2025}{Toronto, ON,
Canada}
\acmBooktitle{Proceedings of the 31st ACM SIGKDD Conference on Knowledge
Discovery and Data Mining V.1 (KDD '25), August 3--7, 2025, Toronto, ON,
Canada}
\acmDOI{10.1145/3690624.3709393}
\acmISBN{979-8-4007-1245-6/25/08}

\begin{document}
\begin{sloppypar}

\title{FuzzyLight: A Robust Two-Stage Fuzzy Approach for Traffic Signal Control Works in Real Cities} 


\author{Mingyuan Li}
\authornotemark[1]
\affiliation{%
  \institution{College of Cyberspace Security, \\ Beijing University of Posts and Telecommunications,}
  \city{Beijing}
  \country{China}
}
\email{henryli_i@bupt.edu.cn}

\author{Jiahao Wang}
\authornote{Contributed equally to this research}
\affiliation{%
  \institution{College of Cyberspace Security, \\ Beijing University of Posts and Telecommunications,}
  \city{Beijing}
  \country{China}
}
\email{jhwang@bupt.edu.cn}

\author{Bo Du}
\affiliation{%
\institution{Department of Business Strategy and Innovation,\\ Griffith University,}
 \city{Brisbane}
 \country{Australia}}
 \email{bo.du@griffith.edu.au}

\author{Jun Shen}
\affiliation{%
  \institution{School of Computing and Information Technology,\\ University of Wollongong,}
  \city{ Wollongong}
  \country{Australia}
  }
  \email{jshen@uow.edu.au}

\author{Qiang Wu}
\authornote{Qiang Wu is the corresponding author}
\affiliation{%
  \institution{Collaborative Innovation Center for Western Ecological Safety,\\ Lanzhou University}
  \city{Lanzhou}
  \country{China}
}
\email{wuq17@lzu.edu.cn}

\begin{abstract}
Effective traffic signal control (TSC) is crucial in mitigating urban congestion and reducing emissions.
Recently, reinforcement learning (RL) has been the research trend for TSC.
However, existing RL   algorithms face several real-world challenges that hinder their practical deployment in TSC:
(1) Sensor accuracy deteriorates with increased sensor detection range, and data transmission is prone to noise, potentially resulting in unsafe TSC decisions.
(2) During the training of online RL, interactions with the environment could be unstable, potentially leading to inappropriate traffic signal phase (TSP) selection and  traffic congestion.
(3) Most current TSC algorithms focus only on TSP decisions, overlooking the critical aspect of phase duration, affecting  safety and efficiency.
To overcome these challenges, we propose a robust two-stage fuzzy approach called FuzzyLight, which integrates compressed sensing and RL for TSC deployment.
FuzzyLight offers several key contributions:
(1) It employs fuzzy logic and compressed sensing to address sensor noise and enhances the efficiency of TSP decisions.
(2) It maintains stable performance during training and combines fuzzy logic with RL to generate precise phases.
(3) It works in real cities across 22 intersections and demonstrates superior performance in both real-world and simulated environments.
Experimental results indicate that FuzzyLight enhances traffic efficiency by 48\% compared to expert-designed timings in the 
 real world. 
Furthermore, it achieves state-of-the-art (SOTA) performance in simulated environments using six real-world datasets with transmission noise.
The code and deployment video are available at the URL\footnote{https://github.com/AdvancedAI-ComplexSystem/SmartCity/tree/main/FuzzyLight}.
\end{abstract}




\maketitle
\newcommand\kddavailabilityurl{https://doi.org/10.1145/3690624.3709393}

\ifdefempty{\kddavailabilityurl}{}{
\begingroup\small\noindent\raggedright\textbf{KDD Availability Link:}\\
The source code of this paper has been made publicly available at \url{\kddavailabilityurl}.
\endgroup
}

\section{Introduction}
\label{Sec:Intro}

\begin{figure}[ht]
\centering
\includegraphics[width=0.45\textwidth]{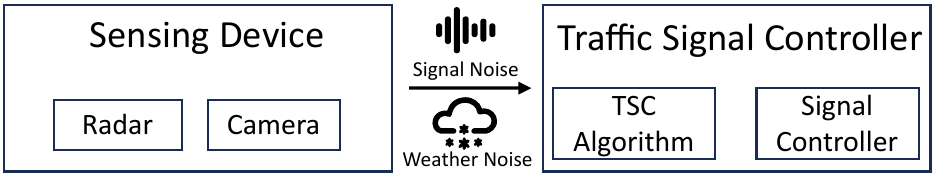}
\caption{Using sensing devices (radar and cameras) to get vehicle data. Then, the data input to the TSC algorithm is susceptible to noise interference.}
\label{fig:system}
\end{figure}
\textbf{Motivations.} 
Real-time adaptive traffic signal control (TSC) enhances urban transportation network efficiency~\cite{wei2019survey,zheng2019frap,wei2019colight,oroojlooy2020attendlight}.
In real-world TSC deployments (as shown in Figure~\ref{fig:system}), existing solutions~\cite{sensor_tsc} utilize terminal sensing devices like radars and cameras to capture real-time traffic data.
These data are then processed and transmitted to the TSC algorithm, which determines the appropriate  traffic signal phase (TSP) or phase duration. 
However, the transmission process is vulnerable to noise interference from various sources, including adverse weather conditions~\cite{weather_sensor} and network factors~\cite{noise_tranmission}. 
Noise-affected data can lead to incorrect TSP or phase duration, potentially causing traffic congestion and safety hazards~\cite{du2022safelight,mei2023reinforcement}.

Most current reinforcement learning (RL) TSC methods rely on online learning, which requires a trial-and-error process with neural networks to determine the appropriate TSP. 
This dependence on real-time interaction during training could result in inefficient phase selections, which exacerbates traffic congestion.
Furthermore, most RL-based methods focus solely on TSP selection without specifying phase duration.
Generating appropriate phase duration is crucial for enhancing driver anticipation, preventing sudden road incidents, and improving TSC safety~\cite{cycle_duration}. 
It also allows better adaptation to current road conditions, thus increasing vehicle throughput~\cite{duration_tsc}.

%

\textbf{Challenges.}
Recent RL advances have enabled TSC algorithms to surpass traditional rule-based methods. However, existing RL-based TSC algorithms face several real-world challenges that hinder their practical deployment: 

(1) Sensor accuracy deteriorates with increased detection range, and data transmission is prone to noise, potentially resulting in poor TSC performance.
Although compressed sensing technology has shown promise in noise suppression through sparse representation and regularization techniques~\cite{cs,l1norm,donoho2006compressed}, it faces limitations in TSC applications. 
Real-time traffic flow data often lacks the necessary sparsity, and sensors have limited detection ranges~\cite{sense_1}. 
Expanding detection areas significantly increases deployment costs. 
Notably, the optimal TSC performance is achieved using vehicle information within a specific range~\cite{wu2021efficient}.

(2) The training process of RL requires extensive interaction with the intersection environment, leading to inefficiencies in training and limited generalization to diverse traffic scenarios. 
Traditional RL-based TSC algorithms often need millions of interactions with the environment to converge~\cite{Wei2019In}, which is impractical for real-world deployment. 
Moreover, these algorithms typically struggle to generalize across different intersection layouts and traffic patterns~\cite{chen2020toward}.
Recent work has attempted to address these issues through transfer learning~\cite{liu2020traffic} and meta-learning~\cite{meta} approaches, but challenges remain in achieving efficient training and robust generalization across a wide range of urban traffic conditions.

(3) Most current TSC algorithms focus exclusively on TSP  decision, overlooking the critical aspect of phase duration. This oversight affects both driver and pedestrian efficiency.
Existing online TSC DRL algorithms primarily output TSP decisions. 
Methods for determining phase duration often rely on preset configurations based on expert experience~\cite{zhang2022expression,liang2019deep,zhao2022ipdalight,zhang2022dynamicLight}, failing to adjust green light durations based on real-time traffic data. 
These approaches can result in underutilized or insufficient green light times, leading to idle periods that reduce traffic efficiency or cause  congestion and delays.

\textbf{Contributions.} 
In this paper, we propose FuzzyLight, a robust two-stage fuzzy approach that integrates compressed sensing and RL for TSC. 
FuzzyLight leverages fuzzy logic~\cite{zadeh1965fuzzy} to handle uncertainty and imprecision in TSC. 
Integrating multiple factors (e.g., traffic flow and expert knowledge) and employing diverse rules and fuzzification  functions offer a comprehensive solution to address various TSC challenges efficiently.
Our main contributions are summarized as following:

\textbf{(1) Noise-resistant TSC with effective sensor range:} 
We combine fuzzy logic with compressed sensing to effectively mitigate transmission noise, improve robustness, and guide optimal sensor deployment.
Moreover, we introduce the concept of an effective sensor detection range for TSC. Within this range, traffic data is transformed into a sparse matrix, enhancing processing efficiency while preserving critical information. 

 \textbf{(2) Two-stage fuzzy approach for efficient and stable TSC:} FuzzyLight employs a novel two-stage process: (a) fuzzy rules for appropriate TSP selection, and (b) a fuzzy function integrated with RL for precise phase duration prediction. 
This approach ensures high performance and stability during training while adapting to real-time traffic flow for enhanced efficiency.

\textbf{(3) Superior performance in real-world and simulated environments:} FuzzyLight demonstrates up to {48\% efficiency improvement} over expert-designed systems across 22 intersections in {two real-world cities}.
In simulated environments using CityFlow, it achieves competitive or state-of-the-art (SOTA) results on multiple datasets, surpassing baseline  algorithms. 
FuzzyLight also exhibits better  generalizability, outperforming solid baseline methods.

\section{Background}
\subsection{Traffic Signal Control} 
A traffic network comprises multiple intersections, each with various lanes facilitating vehicle movement across the intersection, including turning left, proceeding straight, or turning right.

An incoming lane is defined as a lane where vehicles enter an intersection, while an outgoing lane facilitates the exit of vehicles. The set of incoming lanes at intersection $i$ is denoted as $\mathcal{L}^{in}_i$.
\begin{figure}[h]
\includegraphics[width=0.5\textwidth]{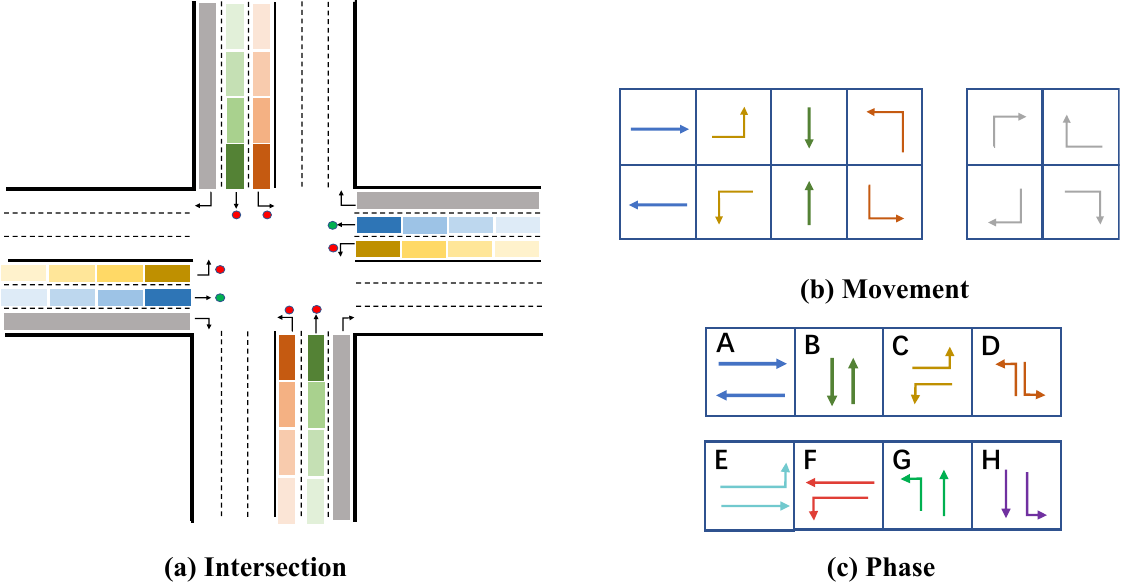}
\caption{Definitions of traffic signal control.}
\label{fig:preliminary}
\end{figure}

\begin{figure*}[ht]
    \centering
    \includegraphics[width=0.9\textwidth ]{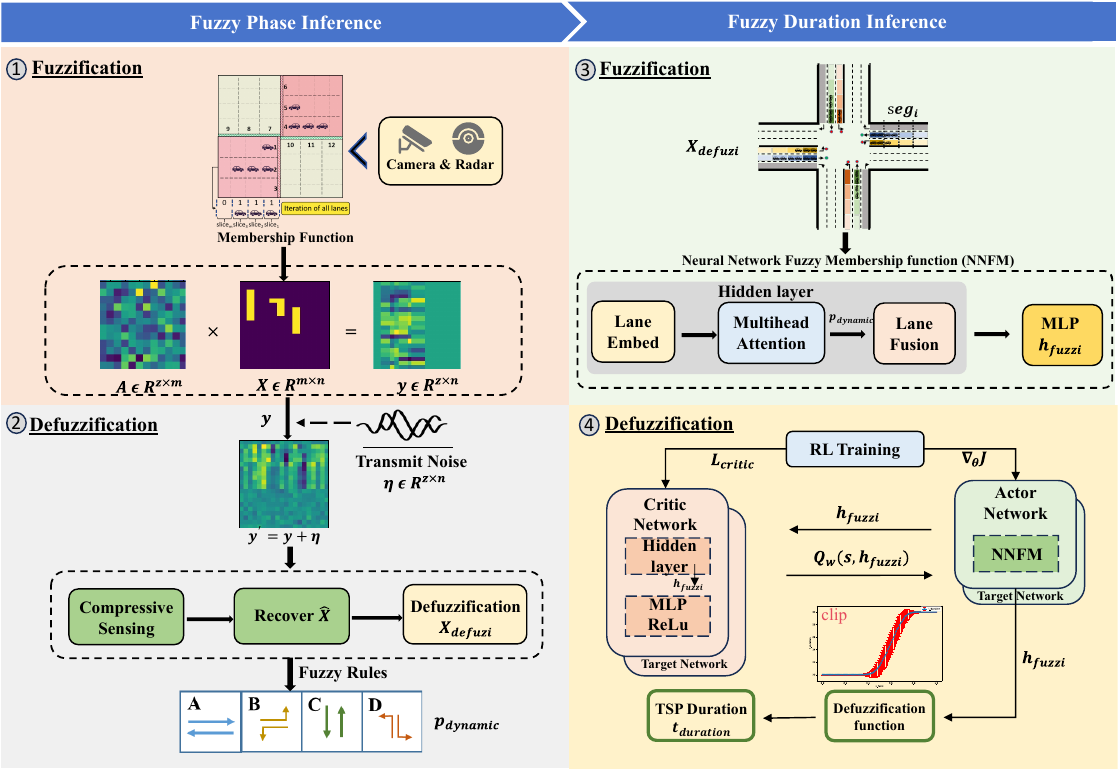}
     \caption{The overall structure of FuzzyLight. FuzzyLight consists of two stages. First, it combines compressive sensing and fuzzy logic to denoise and transmit sensor data, ensuring robustness. Then, it uses fuzzy rule to select the appropriate phase. Second, it uses neural network membership function and defuzzification functions to output the precise phase duration.}
    \label{fig:network}
    \vspace{-1em}
\end{figure*}
Traffic movement refers to vehicles crossing an intersection in a specified direction: left, straight, or right. As illustrated in Figure \ref{fig:preliminary}(b), each intersection typically supports twelve traffic movements, eight pivotal in formulating signal phases.

A traffic signal phase comprises a set of permitted traffic movements. As depicted in Figure \ref{fig:preliminary}(c), we consider $12$ traffic movements, and $p_i$ represnts the  $i-th$ TSP . In this configuration, we use four distinct TSP phases.

The phase duration, denoted by $t_{duration}$, corresponds to the period during which the traffic light remains green. This parameter is critical to optimizing traffic flow at intersections.

\label{ref:pre}
The state representation of each intersection is based on individual lanes, including metrics such as the total number of vehicles per lane ($x(l), l \in \mathcal{L}^{in}_i$), segmentation of each lane defined as $x_i(l)$ with segments spaced $m$ meters apart, and the queue length, which measures the number of waiting vehicles ($q(l), l \in \mathcal{L}^{in}_i$).

\subsection{Fuzzy Logic}
Fuzzy logic is a reasoning method based on fuzzy set theory designed to handle uncertainty and vagueness in information~\cite{czabanski2017introduction,gu2023autonomous}. 
Unlike classical binary logic, fuzzy logic allows elements in a set to have partial membership degrees from membership functions, rather than just zeros or ones.
By employing fuzzy logic, we can better capture the fuzzy and vague relationships and rules in complex systems, making it an effective tool for dealing with uncertainty in real-world environments.

The fuzzyification technique involves transforming input into fuzzy sets $F_i$ and $U_i$, it can be a neural network or any other membership function~\cite{berenji1992learning,berenji1992reinforcement}. Simple inputs refer to the information obtained from the traffic network, such as queue lengths.

Reasoning Engine pertains to the determination of which specific fuzzy rule is to be executed in a given context.
Fuzzy Rules~\cite{fathinezhad2016supervised} contain a set of rules in the form of IF-THEN conditions:
\begin{equation}
    \begin{aligned}    
   For\text{ } rules = 1, 2, ..., N: 
   \\ \text{if }x_1 \in F_1,\text{ } x_2 \in F_2
   \\ \text{Then y is } U_i, i= 1,..,N
  \end{aligned} 
  \notag
\end{equation}
where $x$ and $y$ are inputs and outputs, $F$ and $U$ denote fuzzy sets.

Defuzzification is a step in fuzzy logic that transforms the fuzzy set into accurate output. It performs the defuzzification process in which the membership relations of fuzzy sets are mapped to actual values, converting the fuzzy output into a clear, concrete, and non-fuzzy result for subsequent decision or control operations.

\subsection{Compressed Sensing}
Compressed Sensing~\cite{donoho2006compressed} is an advanced signal processing technique that enables the reconstruction of signals from a small number of measurements.  
The principles of compressed sensing can be represented by the following equation:
\begin{equation}
    \begin{aligned}
    y = Ax+\eta
    \end{aligned}
\end{equation}
where $A \in \mathbb{R}^{m \times n}$ is the observation matrix, $y$ is the observed signal, $x\in \mathbb{R}^{n}$ is the original signal, and $\eta \in \mathbb{R}^{m}$ represents the noise interference. Given the observed signal $y$ and the observation matrix $A$, we aim to recover the original signal $x$ with far fewer observations than required by the Nyquist sampling theorem. To solve this underdetermined problem, we assume that $x$ is sparse. This transforms the optimization problem into the following equation:
\begin{equation}
    \begin{aligned}
    \min _x\|x\|_1 \quad subject \text{ } to \quad\|y-A x\|_2 \leq \delta
    \end{aligned}
   \label{eqation:cs}
\end{equation}
And $\delta$ is the upper limit of noise estimation.
Using the Basis Pursuit Denoising (BPDN) algorithm~\cite{lu2010modified}, we can recover the original signal $\hat{x}$ and allow some error $\delta$ between $\hat{x}$ and $x$. This approach leverages the sparsity of $x$ to find the solution that best fits the observed data $y$ while minimizing the impact of noise $\eta$.

\subsection{Reinforement Learning 
} 
The Markov decision process (MDP) is defined as $M=(S,A,P,r,\gamma) $, where $S$ denotes a set of states,  $A$ is a set of actions, $P$ defines a transition probability, $r$ is a reward function, and $\gamma$ ($\gamma \in [0,1]$) is a discount factor.
    The RL algorithm usually learns a policy $\pi$ online for MDP to determine the best action $a$ ($a \in A$) considering the given state $s$ ($ s\in S$). 
    The objective of RL is to maximize the expected return  $G_t$ at time $t$:
    \begin{equation}
        G_t= \sum_{t=0}^{+\infty}\gamma^{n}r_{t} 
    \end{equation}
 
\section{Methods}
\label{Sec:Methods} 
In this section, we elucidate the structure and functionality of our proposed FuzzyLight model, which is segmented into two distinct stages: Fuzzy Phase Inference and Fuzzy Duration Inference.
These stages are designed to generate the TSP and their respective durations (as shown in Figure~\ref{fig:network}).
%

\subsection{Fuzzy Phase Inference}
The Fuzzy Phase Inference stage employs compressed sensing techniques to process data into a sparse matrix format and subsequently defuzzifies it to extract accurate traffic data (vehicle information). 
The system uses fuzzy logic rules to determine the optimal TSP for the current traffic conditions.
\subsubsection{Fuzzification: membership function with compressed sensing}
Initially, to meet the requirements of compressed sensing, a negotiation between the sensor and the traffic signal controller establishes a random matrix $A \in \mathbb{R}^{z \times n}$. 
Following this, we construct the sensor data into a sparse matrix $X \in \mathbb{R}^{n\times m}$ to fulfill the sparsity condition essential for effective compressed sensing. 
It represents the number of waiting vehicles in all lanes, where $n$ represents the number of entry lanes at the intersection $\mathcal{L}^{in}$, and $m=\frac{ER}{slice}$. 
In this matrix, $slice=v_l+safe_l$, where $v_l$ is the vehicle length, enhancing the sparsity, and $safe_l$ is the safety distance between vehicles. The $ER$ denotes the effective observation range used to calibrate the sensor deployment, optimizing cost and performance.

Next, the distance $d_i$ of each vehicle in lane $i$ from the stop line is measured using sensors (e.g., radar, cameras). These measurements are used to populate the sparse matrix as per the membership functions:
\begin{equation}
X\left[ i,j \right] =\begin{cases}1&if\  \  \  j<\frac{d_{i}}{slice} <j+1,\\ 0&else\end{cases}
\label{sparse}
\end{equation}
The processed data $y = AX$ is transmitted from the sensors to the traffic signal controller. During transmission, noise $\eta \in \mathbb{R}^{z \times m}$ can interfere, resulting in $y'=AX + \eta$.
Compressed sensing is applied  to reconstruct the original data $\hat{X}$ according to Equation~\ref{eqation:cs}. 
\subsubsection{Defuzzification: denoising function and  fuzzy rules for TSP}
%
Subsequently, a defuzzification function is applied to obtain precise vehicle information:
\begin{equation}
X_{defuzi}=round(\max(\hat{X},0))
\label{stage1.defu}
\end{equation}
where $round$ signifies rounding up to the nearest integer. 
Based on $X_{defuzi}$, the number of waiting vehicles for each phase $WV_{p_i}$ is calculated, and fuzzy rules (detailed in Algorithm~\ref{algo:fuzzy phase inference} lines 7-9) are used to determine the phase $p_{dynamic}$ corresponding to the largest number of waiting vehicles.

The specific procedures are detailed further in Algorithm~\ref{algo:fuzzy phase inference}.

\begin{algorithm}
\caption{Fuzzy Phase Inference Algorithm}
{\bf Input:} Vehicle distance $d_i$, incoherence matrix $A \in \mathbb{R}^{z \times n}$ 
 \\
{\bf Output:} Signal phase $p_{dynamic}$
 \begin{algorithmic}[1]
  \label{algo:fuzzy phase inference}
    \STATE Construct sparse matrix $X$ using Eq.~\ref{sparse}.
    \STATE Acquire data $y=AX$.
    \STATE Transmit $y$ and receive noisy data $y'=AX + \eta$.
    \STATE Reconstruct data $\hat{X}$ using compressed sensing (Eq.~\ref{eqation:cs}).
    \STATE Apply defuzzification to get precise vehicle data $X_{defuzi}$ (Eq.~\ref{stage1.defu}).
    \STATE Calculate waiting vehicles $WV_{p_i}$ for each phase from $X_{defuzi}$.
       \IF{$WV_{p_i} \geqslant max(WV_{p_j}), i \neq j$}
        \RETURN $p_{dynamic}=p_i$
       \ENDIF{}
 \end{algorithmic}
\end{algorithm}
\vspace{-1em}
\subsection{Fuzzy Duration Inference}
Fuzzy Duration Inference is also guided by fuzzy logic and RL process to calculate the phase duration.
Here, we design a neural network fuzzy membership function to fuzzify the number of waiting vehicles in segments to the [0,1] interval.
Then, the defuzzification function and RL process are designed  to output the precise phase duration.
\subsubsection{Fuzzification: Neural network fuzzy membership function (NNFM)} 
\textbf{Lane Embedding Layer:}
 First, we extract  the number of vehicles of different segments ($seg=k * slice$) of each traffic movement by recovered vehicle matrix $\hat{X}$. Where $k$ defines the length of the segment within each lane.
%
To allow more capacity in feature extraction, each segment of the lane is embedded from one-dimensional into a higher dimensional latent space using a layer of multi-layer perception (MLP):
\begin{gather}
    h_{embed}(l) = \sigma(\hat{X_e}*W_e+b_e)
\end{gather}
where $\hat{X_e}\in\mathbb{R}^{m \times n}$ is lane embedding matrix and $\hat{X_e}=x_i(l)$.
One row $(x_i,i=1,2,...,m)$ expresses the number of waiting vehicles in the segment of each lane. 
$W_e \in \mathbb{R}^{n \times d}$ and $b_e \in \mathbb{R}^{m \times d}$ are weight matrix and bias vector to learn, respectively. $\sigma$ is the sigmoid function. 

\noindent\textbf{Multihead Attention Layer:} To focus on information about vehicles in different segments. 
A multi-head-attention (MHA) layer is used to understand the impact of different segments:
\begin{equation}
    h_{con}(l) = Concat(h_{embed}(l_j),h_{embed}( l_k)),l_j,l_k \in \mathcal{L}_i^{in} 
\end{equation}
\begin{equation} 
h_{att}(p) = Mean(MHA(h_{con}(l_j),h_{con}(l_j))),l_j \in \mathcal{L}_i^{in} 
\end{equation}
where $l_j, l_k$ are the lanes corresponding to the phase pair, and $\mathcal{L}_i^{in}$ is the set of participating entering lanes. 
%

\noindent\textbf{Lane Fusion Layer:} To learn the segment information of the lanes corresponding to different phase pairs. 
We concat the lane attention information of each phase pair and multiply it with the one hot phase vector:
\begin{equation}
\begin{aligned}
    h_{fusion}(p) = Multiply(p_{onehot},Concat(h_{att} \\(p_1),h_{att}(p_2),..,h_{att}(p_i))), p_i \in \mathcal{P} 
\end{aligned}
\end{equation}
where $p_{onehot} \in \mathbb{R}^{1 \times n}$ is phase output from fuzzy phase inference and $n$ is the number of phase.

\noindent\textbf{Multilayer Perceptron (MLP):} Then, using two linear layers and Relu function to output the embedding hidden layer:
\begin{equation}
\begin{aligned}
      h_{layer_1}(p) = relu(h_{fusion}(p)*W_c + b_c), \\
      h_{final}(p) = relu(h_{layer_1}(p)*W_d + b_d)
\end{aligned}
\end{equation}
Finally, using the sigmoid function to blur the input features into the [0,1] interval:
\begin{equation}
\begin{aligned}
      h_{fuzzi}(p) = sigmoid(h_{final}(p))
\end{aligned}
\end{equation}
\subsubsection{Defuzzification: TSP duration defuzzification function and RL process.}

Next, we design a new defuzzification function to output precise phase duration:
\begin{equation}
t_{duration}=clip\left( h_{fuzzi}\ast refer\  duration+\epsilon ,low,high \right)\\ 
\label{eq.green}
\end{equation}
Where the $OU$ noise~\cite{lillicrap2015continuous} $\epsilon$ is used for enhancing phase duration exploration. The $refer$ $duration$ is calculated by maximum lane length $MaxLaneLength$ divided by maximum lane speed $MaxLaneSpeed$: $refer\  duration=\frac{MaxLaneLength}{MaxLaneSpeed}$. 

%
We clip the phase duration to a maximum and minimum value to guarantee safety.
To ensure pedestrian safety, set the pedestrian crossing time as the minimum duration of green lights $low$.

To output the phase duration according to the real-time traffic flow more accurately, we design the RL module based on the Deep Deterministic Policy Gradient(DDPG) \cite{lillicrap2015continuous} to update the fuzzy neural network $\mu_{\theta}$.
The actor and critic networks have the same hidden layers, but the final MLP layer differs between them.
The actor network parameters are updated by the following  equation, $N$ is the sample size:
\begin{equation}
\nabla_{\theta}J \approx \frac{1}{N} \sum_{i=1}^{N}(\nabla_{\theta}\mu_{\theta}(s_i)Q_w(s_i,h_{fuzzi}) |_{h_{fuzzi}=\mu_{\theta}(s_i)}).
 \label{eq1}
\end{equation} 

The critic network parameters are updated by the following equation, $s_{i+1}$ is the next state observed from the environment:
\begin{equation}
\begin{aligned}
      L_i=r_i + \gamma Q_{w^-}(s_{i+1}, \mu_{\theta^-}(s_{i+1})), \\
      L_{critic} = \frac{1}{N}\sum_{i=1}^{N}(L_i - Q_w(s_i, h_{fuzzi}^i))^2 
\label{eq2}
\end{aligned}
\end{equation}

Both the actor and critic networks are updated by  function:
\begin{gather}
      w^- = \tau w + (1-\tau)w^-\notag
\\\theta ^-=\tau\theta + (1-\tau)\theta^-
\label{eq3}
\end{gather}
where $w^-$ and $\theta^-$ are parameters of critic and actor target networks, respectively.

 The detailed algorithm is shown in Algorithm~\ref{algo:algorithm-3}.
 \begin{algorithm}
\caption{Fuzzy duration inference}
\label{algo:algorithm-3}
{\bf Initialize:} Actor network $\mu_{\theta}$, Critic network $\mathcal{Q}_{w}$, OU noise $\epsilon$ for exploration, recover data $\hat{X}$, Replay buffer $\mathcal{R}$.

{\bf Output:} Target actor and critic network $\mu_{\theta^-}, \mathcal{Q}_{w^-}$, real-time phase duration $t_{duration}$: $\theta^- \leftarrow \theta$, $w^- \leftarrow w$
\begin{algorithmic}[1]
\WHILE{Each step in environment}
  \STATE Get state $s$ by $\hat{X}$ from each intersection.
  \STATE Fuzzify $s$ to $h_{fuzzi}$ by NNFM.
  \STATE Get ($s,h_{fuzzi},r,{s}'$) put it into buffer $\mathcal{R}$.

  \FOR{Each episode in $epochs$}
  \STATE Sample data from replay buffer $\mathcal{R}$.
  \STATE Update critic network by Eq.(\ref{eq2}). 
  \STATE Update actor network (NNFM) by Eq.(\ref{eq1}).
  \STATE Update target network by Eq.(\ref{eq3}).
  \ENDFOR
   \STATE Defuzzify $s$ and get phase time $t_{duration}$ by Eq.(\ref{eq.green}).
  \ENDWHILE
\end{algorithmic}
\end{algorithm}

The state, action, and reward are defined below:
\begin{itemize}
\item[$\bullet$]\textbf{State.}
    The state is the number of waiting vehicles in each segment $s=x_i(l)$. 
    \item[$\bullet$]\textbf{Action.} The action is $h_{fuzzi}$ for Eq.(\ref{eq.green}).
    \item[$\bullet$]\textbf{Reward.} Negative queue length $r_i=\sum{-q(l)}, l \in \mathcal{L}^{in}_i  $ is denoted as the reward.
\end{itemize}

Detailed structures and hyperparameters of the FuzzyLight are shown in \textbf{Appendix ~\ref{Details of the network}}.

\begin{figure*}[!ht]
 \centering
 
  \subfigure[Throughput of City1]{
         \centering
         \includegraphics[width=0.23\textwidth, height=2.4cm]{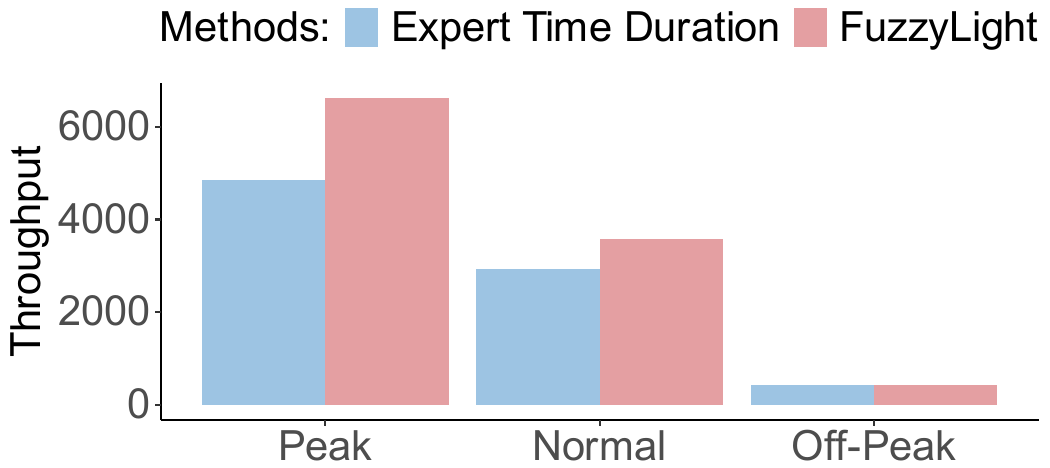}
     }
     \subfigure[Throughput of City2]{
         \centering
         \includegraphics[width=0.23\textwidth, height=2.4cm]{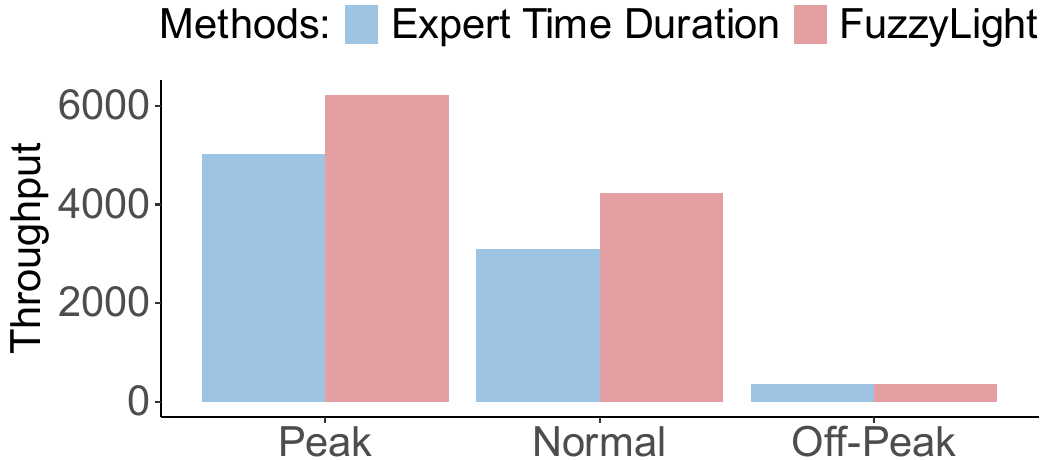}
     }
     \subfigure[Number of Stops for City1]{
         \centering
         \includegraphics[width=0.23\textwidth, height=2.4cm]{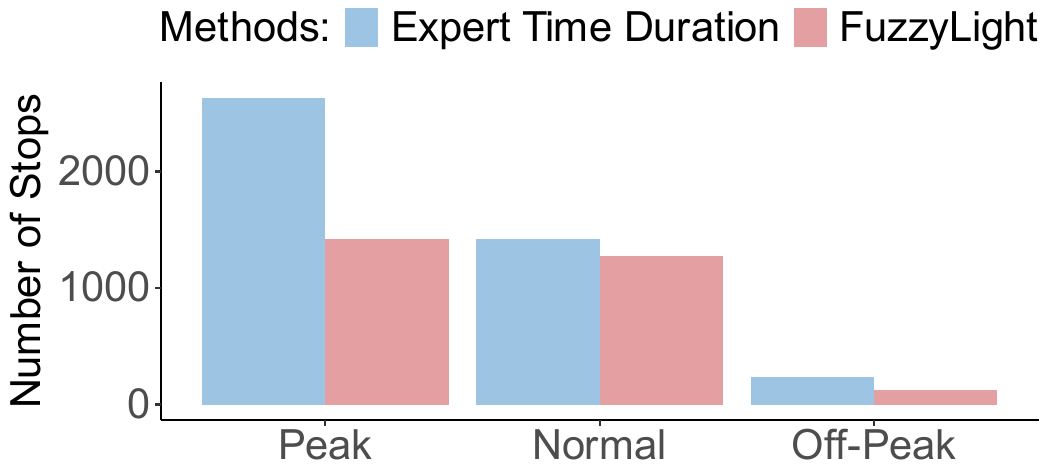}
     }
    \subfigure[Number of Stops for City2]{
         \centering
         \includegraphics[width=0.23\textwidth, height=2.4cm]{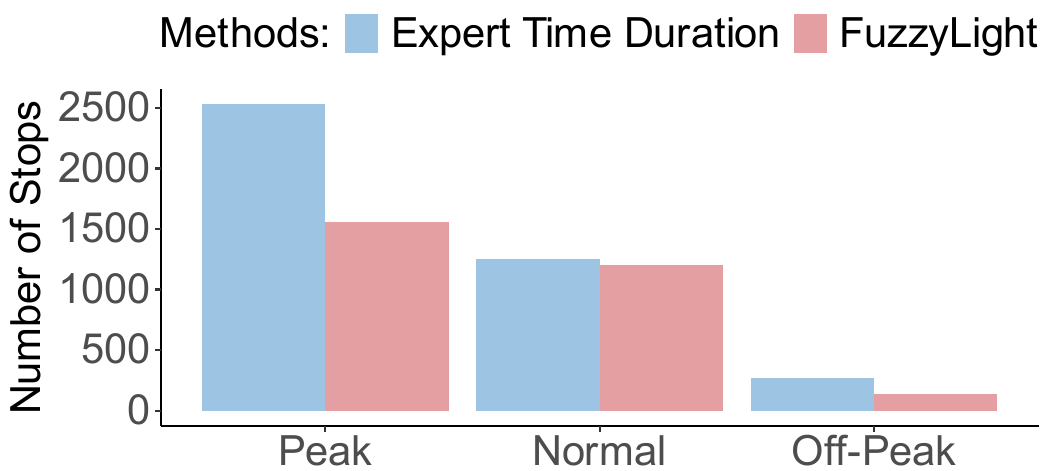}
     }
    \caption{Performance in the real world across two cities during three different time periods (peak, normal, off-peak).}
    \label{fig:realworld_result}
\end{figure*}
\begin{table*}[!ht]
\centering
\caption{Performance of all methods attacked by guassian noise (with max scale) in JiNan, HangZhou in terms of ATT. From top to bottom, the methods are conventional,  RL-based for TSP and phase duration and FuzzyLight. The shorter the ATT, the better.}
\label{table:robust}
\scalebox{1}{
\begin{tabular}{lcccccc} 
\toprule
\multicolumn{1}{c}{\multirow{2}{*}{\textbf{Method}}} & \multicolumn{3}{c}{\textbf{JiNan}}                                        & \multicolumn{2}{c}{\textbf{HangZhou}}           & \textbf{New York}       \\ 
\cline{2-7}
\multicolumn{1}{c}{}                                 & \textbf{1}             & \textbf{2}              & \textbf{3}             & \textbf{1}             & \textbf{2}             & \textbf{1}              \\ 
\hline
\textbf{FixedTime}                                   & 428.11 ± 0.00          & 368.76 ± 0.00           & 383.01 ± 0.00          & 495.57 ± 0.00          & 406.65 ± 0.00          & 1507.12 ± 0.00          \\
\textbf{MaxPressure}                                 & 312.00~± 0.00          & 298.23 ± 0.00           & 290.24~± 0.00          & 355.55~± 0.00          & 368.87~± 0.00          & 1274.50 ± 0.00          \\
\textbf{Efficient-MP}                                & 321.40~± 0.00          & 298.09~± 0.00           & 293.47~± 0.00          & 353.88~± 0.00          & 367.63~± 0.00          & 1254.43~± 0.00          \\ 
\hline
\textbf{MPLight}                                     & 403.73 ±~

42.35       & 501.67 ± 107.24         & 385.49 ± 17.74         & 554.31 ± 83.64         & 396.26 ± 10.75         & 1476.65 ± 112.48        \\
\textbf{CoLight}                                     & 353.76 ±~

11.16       & 324.3 ± 11.28           & 334.9 ± 28.59          & 369.58 ± 11.24         & 382.35 ± 13.79         & 1225.41 ± 139.16        \\
\textbf{Efficient-MPLight}                           & 498.75 ±~

84.15       & 359.59 ± 10.23          & 526.76 ± 17.46         & 454.86 ± 58.61         & 388.51 ± 10.41         & 1739.95 ± 340.3         \\
\textbf{Efficient-CoLight}                           & 365.24 ±~ 16.89        & 329.76 ±~14.47        & 345.69 ± 16.43           & 399.85 ± 29.13           & 382.93 ± 10.03           & 1597.99 ± 153.87        \\
\textbf{DuaLight}                                    & 321.14 ± 9.14          & 301.54 ± 8.14           & 289.41 ± 9.41           & 385.13 ± 21.96         & 361.51 ± 31.17         & 1477.31 ± 216.31        \\
\textbf{Advanced-MP}                                 & 320.99~± 0.00          & 298.28~± 0.00           & 289.37~± 0.00          & 351.60~± 0.00          & 363.70~± 0.00          & 1219.24~± 0.00          \\
\textbf{Advanced-MPLight}                            & 404.34 ± 22.94         & 358.37 ± 29.82          & 373.53 ± 41.09         & 422.33 ± 34.46         & 382.89 ± 34.06         & 1743.19 ± 99.84         \\
\textbf{Advanced-CoLight}                            & 379.06 ± 17.6          & 342.89 ± 20.93          & 322.06 ± 6.33          & 390.05 ± 18.09         & 411.41 ± 19.85         & 1548.34 ± 193.47        \\ 
\hline
\textbf{\textbf{MPCycleTime}}                        & 313.79~± 0.00          & 303.86~± 0.00           & 292.56~± 0.00          & 326.62~± 0.00          & 351.29~± 0.00          & 1286.50 ± 0.00          \\
\textbf{\textbf{\textbf{\textbf{DynamicLight}}}}     & 355.38 ± 17.66         & 369.34 ± 19.01          & 325.17 ± 14.96         & 397.07 ± 8.01          & 389.61 ± 10.75         & 1624.73 ± 39.34         \\
\textbf{\textbf{FuzzyLight}}                         & \textbf{264.39 ± 5.65} & \textbf{251.38 ±~ 6.32} & \textbf{254.86 ± 8.87} & \textbf{284.78 ± 1.29} & \textbf{303.56 ± 2.57} & \textbf{1129.84 ± 9.2}  \\
\bottomrule
\end{tabular}
}
\end{table*}

\section{Experiments}
\label{Sec:Exp}
We conduct extensive experiments in real-world and simulated environments.
Moreover, we demonstrate the performance of FuzzyLight on a scalable and generalizable platform for city-level TSC and heterogeneous intersections.

\subsection{Environments}
We have two types of environments for experiments: real-world environment and simulated environment.\\

\textbf{Real-world environment:}
\begin{itemize}
    \item [$\bullet$]\textbf{City1 dataset:}
    The road network has 4 (2 x 2) intersections. Each intersection is four-way, with two 400-meter (East-West) and two 800-meter (South-North) roads. 
    \item [$\bullet$]\textbf{City2 dataset:}
    The road network has 18 (1 x 14 and 1 x 4) intersections. Each intersection has three lanes in the north-south direction. In the east-west direction, the number of lanes is uneven, with some having 2 or 3 lanes. The lane lengths vary, being either 400 or 800 meters. The intersections are either four-way or three-way junctions.
\end{itemize}

\noindent\textbf{Simulated environment:}
\begin{itemize}
    \item [$\bullet$]\textbf{JiNan datasets:}
    The road network has 12 (3 × 4) intersections. Each intersection is four-way, with two 400-meter (East-West) roads  and two 800-meter (South-North) roads . 
    %
    \item [$\bullet$]\textbf{HangZhou datasets:} The road network has 16 (4 × 4) intersections. Each intersection is four-way, with two 800-meter (East-West) roads and two 600-meter (South-North) roads. 
    %
     \item [$\bullet$]\textbf{New York datasets:} There are 196 (28 × 7) intersections of Manhattan with open-source taxi trip data. 
     %
\end{itemize}

\subsection{Evaluation Metrics and Settings}
We use: (1) the number of stops and throughput as metrics for real-world deployment; and (2) average travel time (ATT)~\cite{chen2022constructing,fang2022monitorlight} as evaluation metrics for the simulated environment. 

Each green signal is followed by a three-second yellow signal and
a two-second all-red time to clear the vehicles at an intersection.
More details on experiment settings in \textbf{Appendix~\ref{app:settings}}.

\subsection{Compared Methods}
For real-world deployment, we use the current expert time duration as the compared method. We conduct a three-month experiments, calculating the number of stops and throughput.

For simulated environment, we use classic traditional and latest DRL methods as the baselines. 
We train these DRL Methods 50 rounds for each model and each round is a 60-minute simulation. 
We compute results as the average and variance of the last ten testing episodes. A brief introduction to these methods is shown below. 

\textbf{Traditional Methods} include Expert time duration (different expert time duration for different traffic flow), FixedTime \cite{koonce2008traffic}, Max-Pressure \cite{Varaiya2013}, Efficient-MP \cite{wu2021efficient}, MPCycleTime\cite{liang2019deep} (Periodic plus or minus 5 seconds) and Advanced-MP \cite{zhang2022expression}.

\textbf{Reinforcement Learning Methods} include MPLight~\cite{chen2020toward}, CoLight~\cite{wei2019colight}, Efficient-MPLight~\cite{wu2021efficient}, Efficient-CoLight~\cite{wu2021efficient}, DualLight~\cite{lu2023dualight}, Advanced-MPLight~\cite{zhang2022expression}, Advanced-CoLight~\cite{zhang2022expression}, and DynamicLight~\cite{zhang2022dynamicLight} (for phase duration).

\subsection{Results}
In this section, we analyze the results of the FuzzyLight on different real-world and simulation environments.

\begin{table*}[ht]
\centering
\caption{Performance without noise for all methods in JiNan, HangZhou regarding ATT. From top to bottom, the methods are conventional, RL-based for TSP and phase duration, and FuzzyLight. The shorter the ATT, the better it is. }
\label{table:1}
\scalebox{1}{
\begin{tabular}{lcccccc} 
\toprule
\multicolumn{1}{c}{\multirow{2}{*}{\textbf{Method}}} & \multicolumn{3}{c}{\textbf{JiNan}}                                                & \multicolumn{2}{c}{\textbf{HangZhou}}                             & \textbf{New York}        \\ 
\cline{2-7}
\multicolumn{1}{c}{}                                 & \textbf{1}             & \textbf{2}                      & \textbf{3}             & \textbf{1}                      & \textbf{2}                      & \textbf{1}               \\ 
\hline
\textbf{FixedTime}                                   & 428.11 ± 0.00          & 368.76 ± 0.00                   & 383.01 ± 0.00          & 495.57 ± 0.00                   & 406.65 ± 0.00                   & 1507.12 ± 0.00           \\
\textbf{MaxPressure}                                 & 273.96 ± 0.00          & 245.38 ± 0.00                   & 243.80 ± 0.00          & 288.54 ± 0.00                   & 348.98 ± 0.00                   & 1179.55 ± 0.00           \\
\textbf{Efficient-MP}                                & 269.87 ± 0.00          & 240.02 ± 0.00                   & 239.75 ± 0.00          & 284.44 ± 0.00                   & 327.62 ± 0.00                   & 1122.00 ± 0.00           \\ 
\hline
\textbf{MPLight}                                     & 300.25 ± 5.23          & 421.01 ± 68.09                  & 507.53 ± 158.09        & 309.35 ± 2.85                   & 346.32 ± 2.39                   & 1890.42 ± 128.07         \\
\textbf{CoLight}                                     & 281.58 ± 2.72          & 257.13 ± 1.71                   & 261.34 ± 4.4           & 301.70 ± 1.04                   & 339.26 ± 5.19                   & 1699.63 ± 285.68         \\
\textbf{Efficient-MPLight}                           & 271.24 ± 2.67          & 247.89 ± 2.29                   & 242.77 ± 1.92          & 286.22 ± 0.99                   & 321.38 ± 4.99                   & 1925.8 ± 218.21          \\
\textbf{Efficient-CoLight}                           & 296.78 ± 6.69          & 244.03 ± 0.75                   & 249.69 ± 3.82          & 285.16 ± 1.68                   & 327.89 ± 3.70                   & 1900.3 ± 93.13           \\
\textbf{DuaLight}                                    & 265.04±1.21            & 242.83 ± 0.61                     & 239.89 ± 1.37            & 286.57 ± 1.31                     & 316.85 ± 3.19                     & 1436.15 ± 39.81            \\
\textbf{Advanced-MP}                                 & 253.61 ± 0.00          & 238.62 ± 0.00                   & 235.21 ± 0.00          & 279.47 ± 0.00                   & 318.67 ± 0.00                   & 1060.41 ± 0.00  \\
\textbf{Advanced-MPLight}                            & 293.75 ± 19.27         & 237.51 ± 1.23                   & 263.03 ± 49.71         & 278.55 ± 1.22                   & 322.58 ± 4.61                   & 2013.8 ± 202.18          \\
\textbf{Advanced-CoLight}                            & 255.3 ± 2.03           & 236.13 ± 0.63                   & 232.76 ± 0.69          & 278.11 ± 1.01                   & 319.39 ± 10.58                  & \textbf{1010.33 ± 43.56}          \\ 
\hline
\textbf{\textbf{MPCycleTime}}                        & 264.08~± 0.00          & 241.74 ± 0.00                   & 240.02~± 0.00          & 282.87~± 0.00                   & 321.80~± 0.00                   & 1177.81~± 0.00           \\
\textbf{\textbf{\textbf{\textbf{DynamicLight}}}}     & \textbf{251.72 ± 0.83} & 236.31 ± 0.91                   & \textbf{231.52 ± 1.31} & 279.69 ± 2.48                   & 310.71 ± 1.58                   & 1117.87 ± 42.94          \\
\textbf{\textbf{FuzzyLight}}                         & 254.48 ± 1.06          & \textbf{\textbf{235.46 ± 0.94}} & 234.62 ± 0.71          & \textbf{\textbf{277.49 ± 0.92}} & \textbf{\textbf{297.51 ± 2.77}} & 1134.08 ± 12.55          \\
\bottomrule
\end{tabular}
}
\end{table*}
\begin{figure*}[ht] 
\begin{minipage}[t]{0.33\linewidth} 
\centering
\includegraphics[width=2in, height=1.6in]{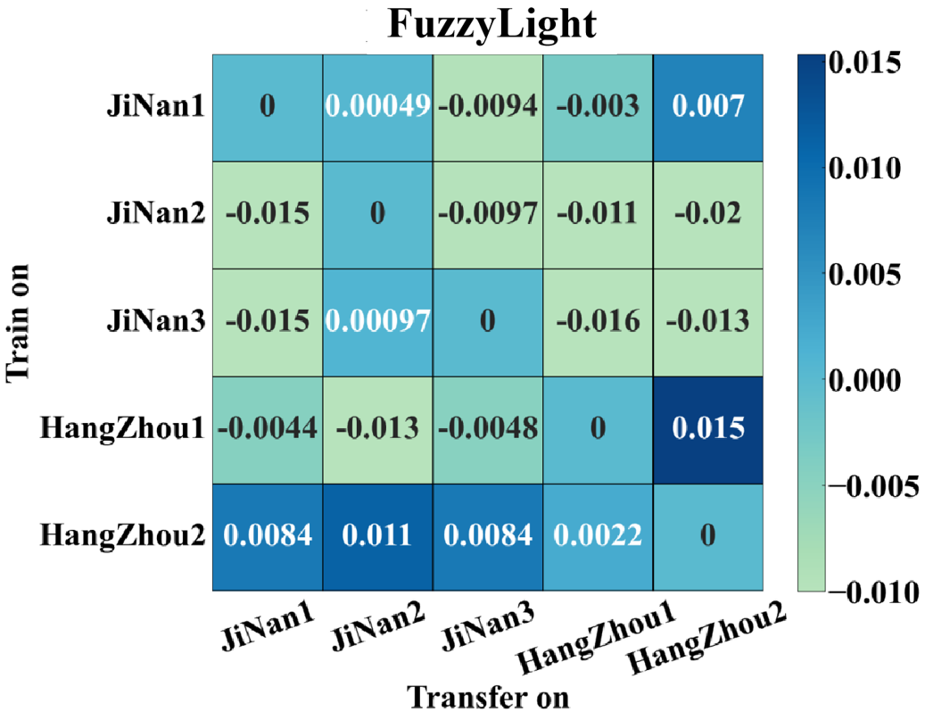} 
\end{minipage}%
\begin{minipage}[t]{0.33\linewidth}
\centering
\includegraphics[width=2in, height=1.6in]{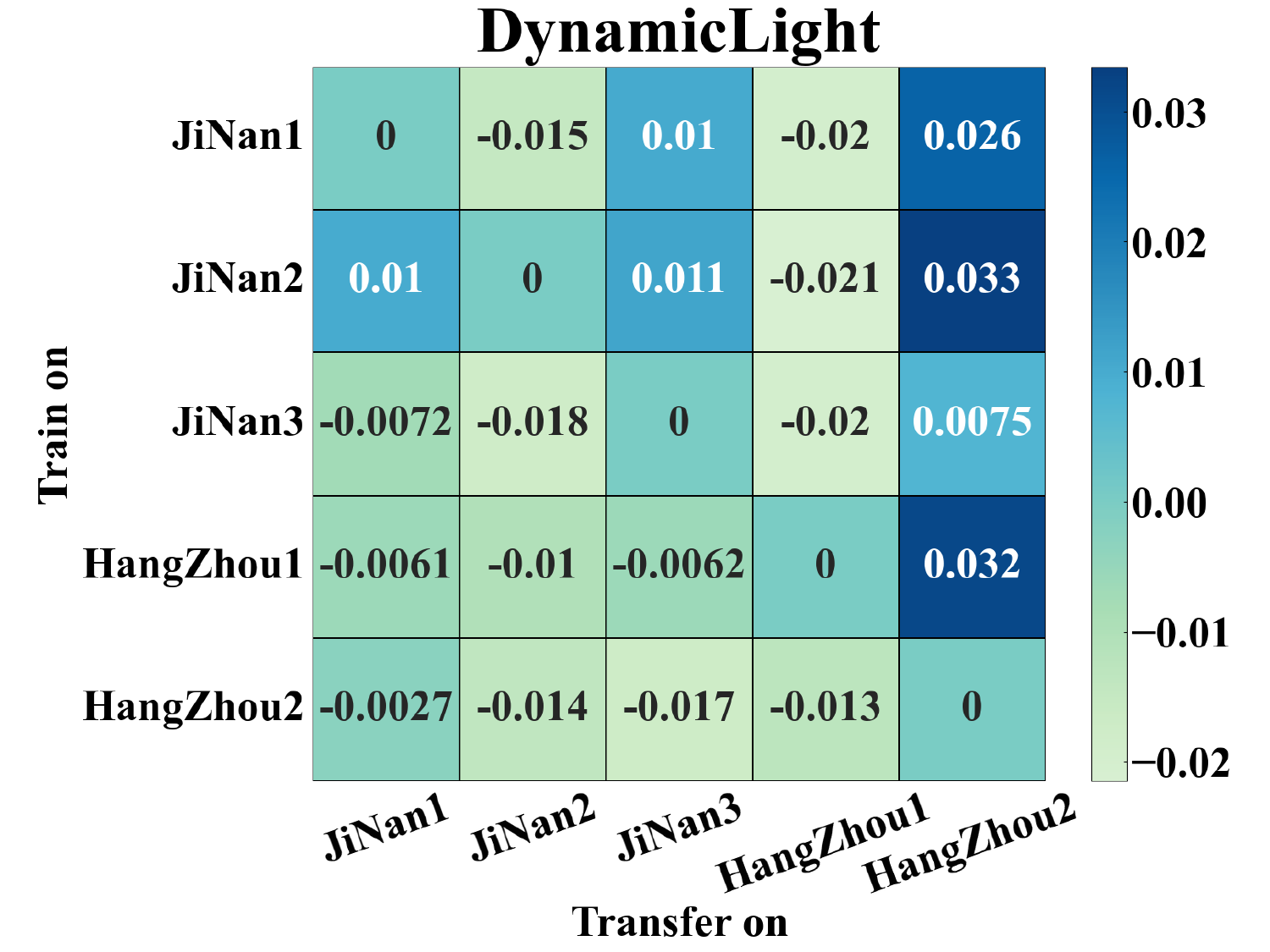}
\end{minipage}%
\begin{minipage}[t]{0.33\linewidth}
\centering
\includegraphics[width=2in, height=1.6in]{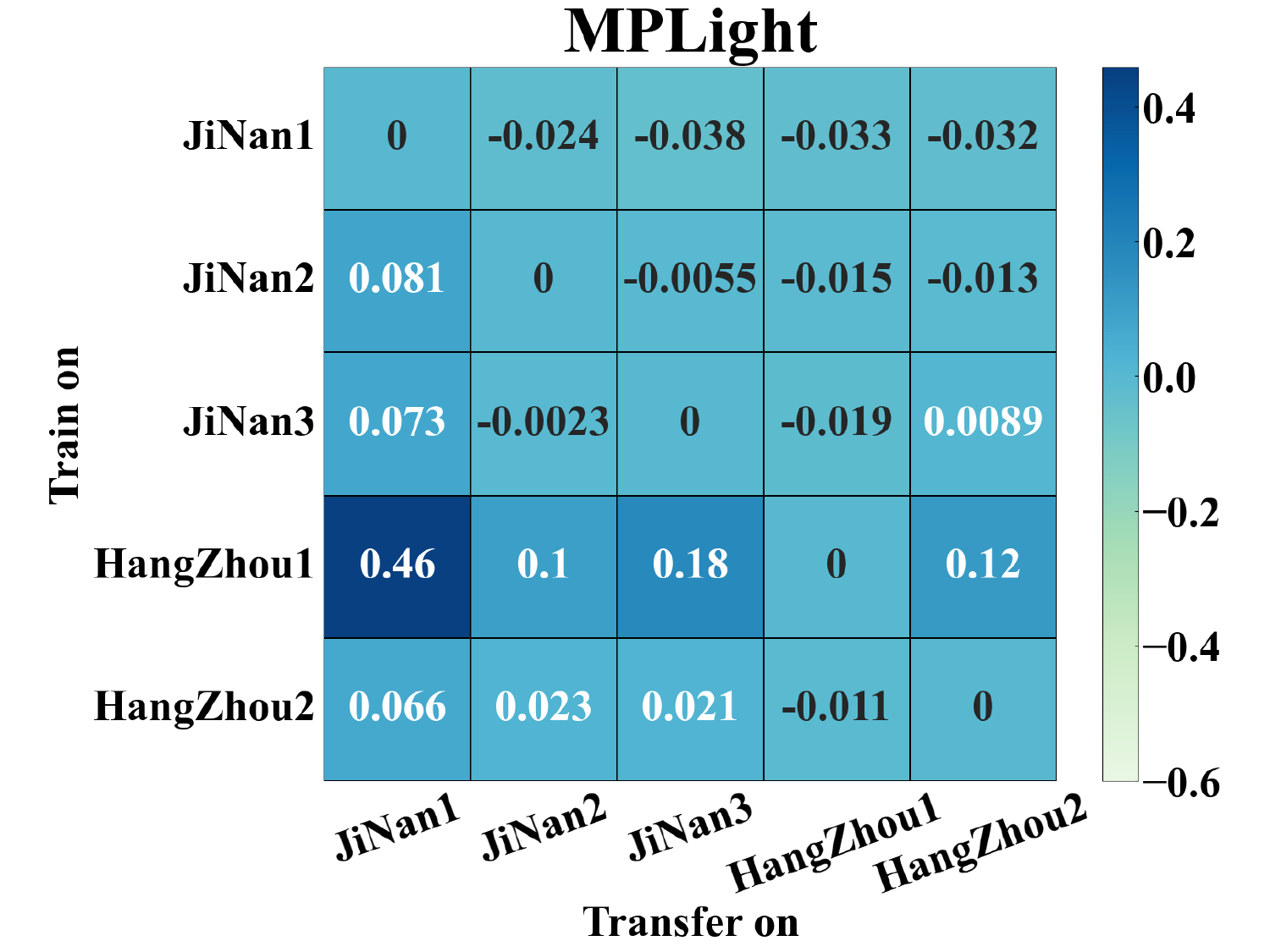}
\end{minipage}
\caption{The results of transfer generalization experiments.} 
\label{fig:transfer} 
\end{figure*}

\subsubsection{Real-world results}
We conducted a three-month experiment with FuzzyLight in the real world, spanning two cities (City1 and City2) and covering 22 intersections. 
Compared to the expert-determined time durations over one month, the results are shown in Figure~\ref{fig:realworld_result}..
We collect data on the average number of vehicle stops and throughput for all intersections during three different time periods: peak hours (including morning and evening peak hours for three hours), normal hours (daytime for six hours), and off-peak hours (other times). The results show that during peak hours, the throughput increases by 36\% in City1 and 24\% in City2, and the number of stops decreases by 46\% in City1 and 38\% in City2. 

During normal hours, the throughput increases by 22\% in City1 and 37\% in City2, and the number of stops decreases by 11\% in City1 and 4\% in City2. During off-peak hours, the throughput increases by 3\% in City1 and 5\% in City2, and the number of stops decreases by 48\% in City1 and 50\% in City2. These experiment results indicate that using our proposed FuzzyLight significantly improves traffic efficiency in both cities across different time periods compared to expert time duration.

The significant improvement during peak hours suggests that FuzzyLight effectively reduces congestion and improves throughput in high traffic volumes by optimizing the phase and duration of the signal. This could be due to its ability to adapt dynamically to fluctuating traffic conditions and prioritize movements that alleviate congestion. While improvements during normal and off-peak hours are also notable, the smaller relative gains compared to peak hours indicate that FuzzyLight is particularly beneficial when traffic demand is high and complex.
During less busy times, the algorithm still enhances efficiency, but the impact is naturally less dramatic due to lower initial congestion levels.
\begin{figure}[h]
        \centering \includegraphics[width=0.9\linewidth]{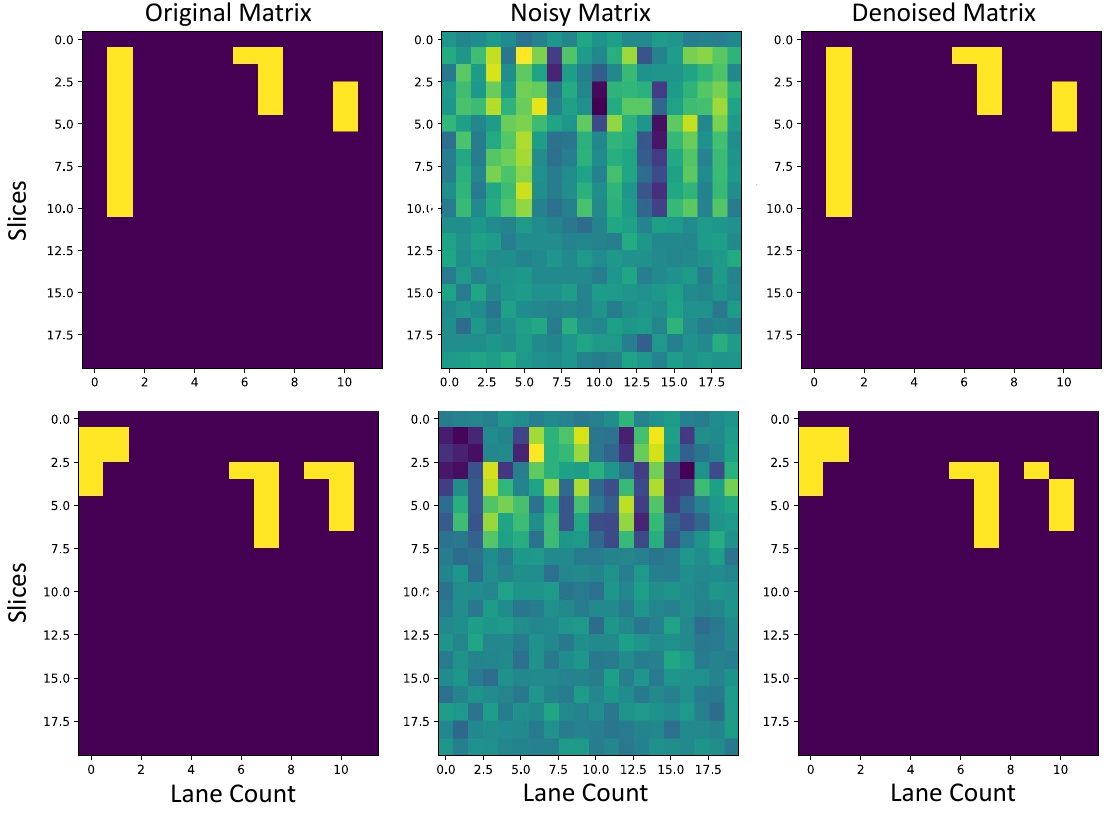}
    \caption{Recovery visualisation $\hat{X}$ in HangZhou1 and JiNan1.}
    \label{fig:recover}
    \vspace{-1em}
\end{figure}

\subsubsection{Simulated  results} We simulate the performance of FuzzyLight in two conditions: the presence of transmission noise and noise free.

\textbf{Performance with noise:} 
As shown in Table~\ref{table:robust}, FuzzyLight outperforms all other methods even under the maximum noise scale and performs close to noise-free performance. Additionally, the visualization of compressed sensing recovery as shown in Figure~\ref{fig:recover} demonstrates that our algorithm effectively recovers data from noise interference and validates its robustness.

\textbf{Performance without noise: }
In the noise free of transmission, as shown in Table~\ref{table:1}, FuzzyLight outperforms other methods across three datasets, achieving state-of-the-art (SOTA) performance. 
Additionally, its performance on other datasets is close to existing SOTA methods, demonstrating that FuzzyLight effectively reduces the average travel time of vehicles.

\subsubsection{Training performance}
\begin{figure}[ht]
 \centering
  \subfigure[Traning result comparison of ATT in JiNan1 with 50 rounds.]{
         \centering
         \includegraphics[width=0.22\textwidth]{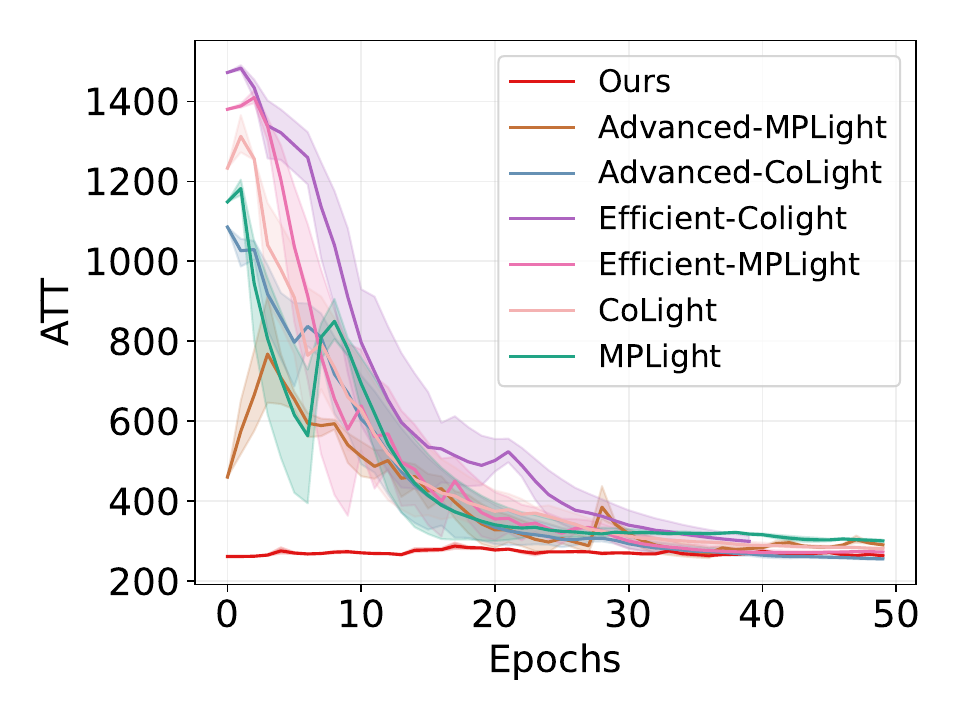}
     }
     \subfigure[Traning result comparison of ATT in HangZhou1 with 50 rounds.]{\centering\includegraphics[width=0.22\textwidth]{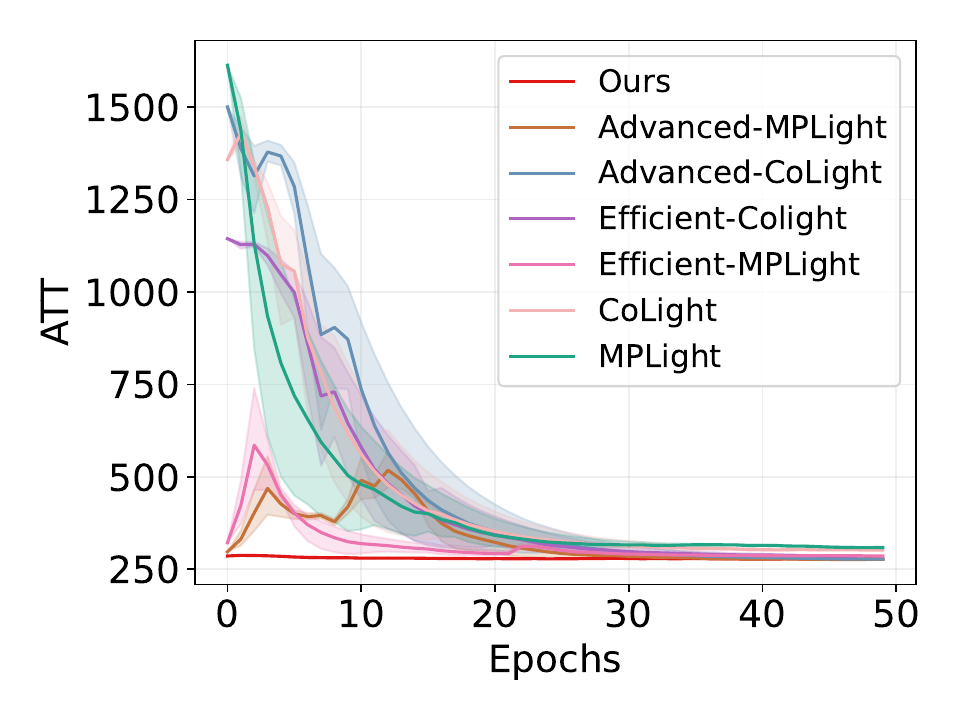}
     }
\caption{The training process of FuzzyLight}
    \label{trainning}
     \vspace{-1.5em}
\end{figure}

We examine the training stability of FuzzyLight to validate if it could deploy to real world environment.
As shown in Figure~\ref{trainning}, FuzzyLight maintains high performance from the early stages of training due to the selection of appropriate phases in phase inference and the constraints on maximum and minimum green light duration in duration inference.
Other RL-based phase selection methods cannot be applied in the real world due to the increased traffic congestion risk during their early trial-and-error training phase. This result guide our real-world deployment.

\begin{figure*}[!ht]
 \centering
  \subfigure[Refer duration analysis]{
         \centering
         \includegraphics[width=0.3\textwidth]{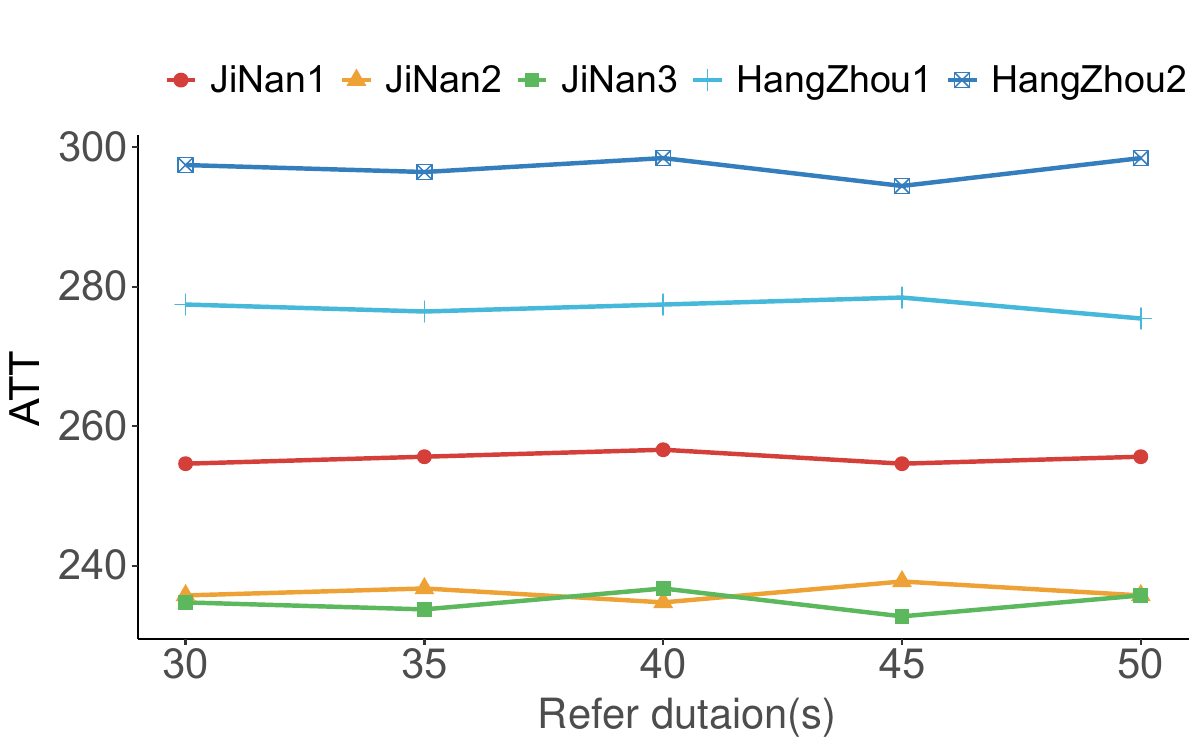}
         \label{fig:refer}
     }
     \subfigure[FuzzyLight based on lane vehicle count]{
         \centering\includegraphics[width=0.3\textwidth]{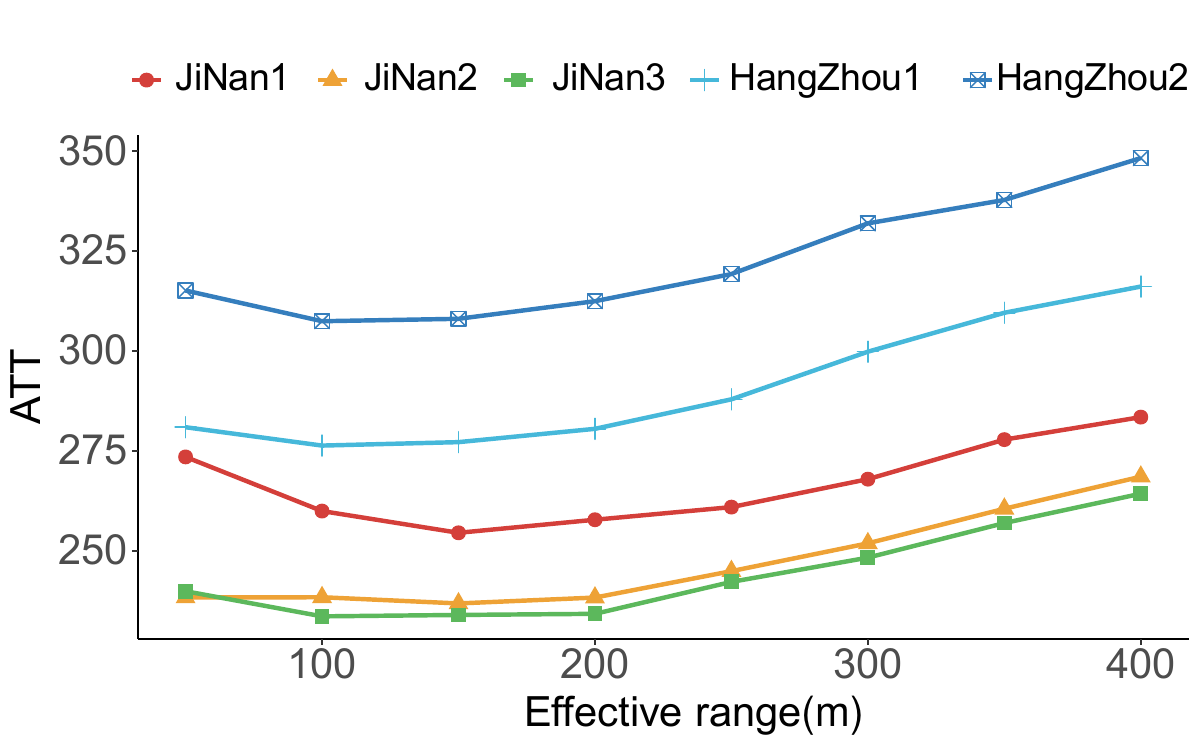}
        \label{er1}
       
     }
     \subfigure[FuzzyLight based on lane queue count]{
         \centering\includegraphics[width=0.3\textwidth]{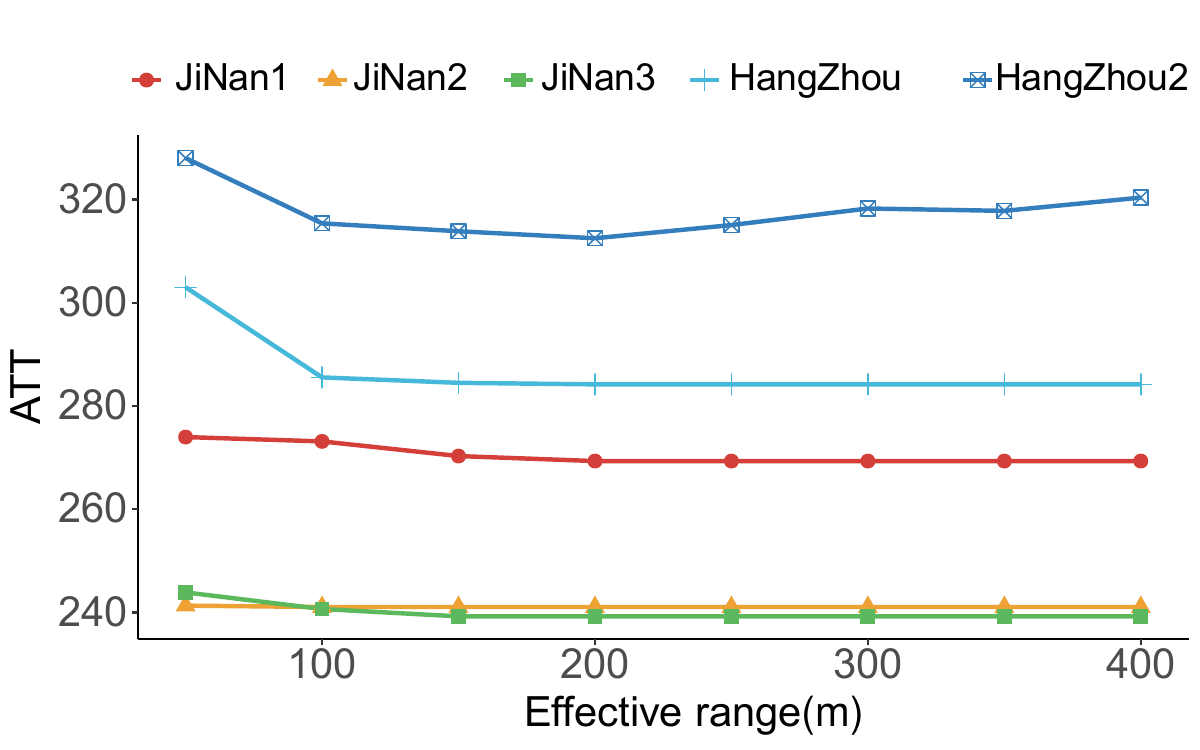}
         \label{er2}
       
     }
    
    \caption{The refer duration and effective range analysis.}
\end{figure*}
\subsubsection{Generalization performance}
For the TSC problem, model generalization is crucial. We transfer a model trained on one dataset to another dataset to achieve better performance. We conduct transfer experiments on five datasets to verify the model's generalization ability. The following metric is used to evaluate the model's transferability: $v_{transfer} = \frac{t_{transfer}}{t_{train}} - 1$. The transfer results are shown in Figure~\ref{fig:transfer}. For $v_{transfer}$, a smaller value is better. We observe that the JiNan and HangZhou datasets exhibit high transferability.

\subsection{Ablation Experiments}
\begin{figure}[H]
 \centering
   \subfigure[Ablation study for duration inference.]{
         \centering
         \includegraphics[width=0.22\textwidth]{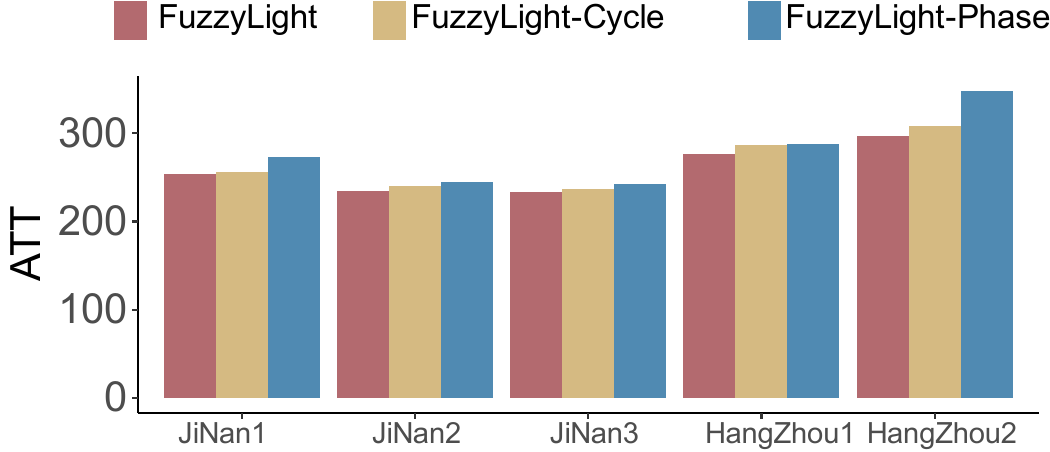}
         \label{fig:stage}
     }
\subfigure[Ablation study for using fuzzy logical.]{
         \centering\includegraphics[width=0.22\textwidth]{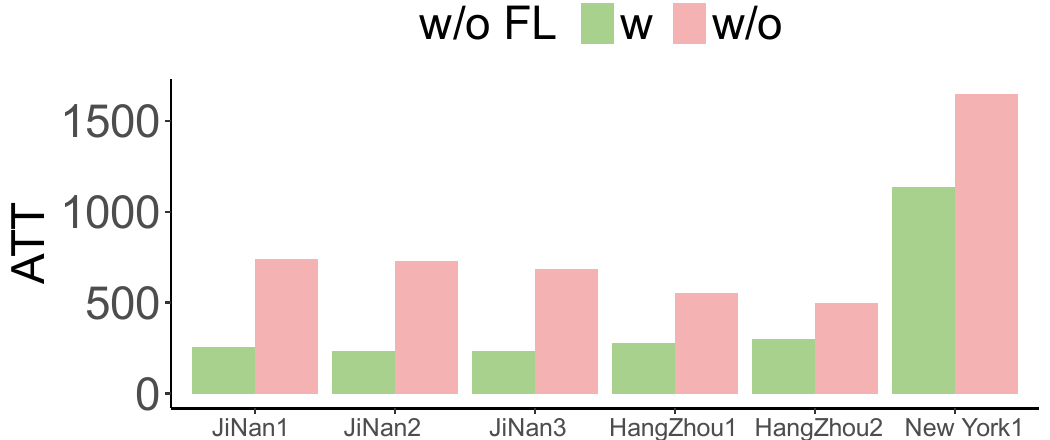}
       \label{wocs}
     }
    \caption{Ablation experiments.}
\end{figure}

To verify the effectiveness of the fuzzy phase and fuzzy duration inference, we conduct comparison experiments. We use FuzzyLight-Cycle (cycle phase selection and duration inference) and FuzzyLight-Phase (fuzzy rules phase selection and fixed duration) for comparison, the result is shown in Figure \ref{fig:stage}.
After using both fuzzy phase and fuzzy duration inference, the ATT improves about 3. 7\% compared to FuzzyLight-Cycle and FuzzyLight-Phase.
It demonstrates the effectiveness of the two-stage process.

To validate the effectiveness of the fuzzy logical in compressed sensing, we conduct a comparison without using fuzzy logical (w/o). As shown in Figure~\ref{wocs}, incorporating the fuzzy logical results in significant performance improvement under transmission noise across five datasets and it confirms the effectiveness of FuzzyLight.
More experiments are provided in \textbf{Appendix \ref{app:exp}}.

\section{Discussion}
\textbf{(1) Why not use  other SOTA RL algorithms in the real world?} 

The deployment of current SOTA RL algorithms in real-world scenarios is hindered by issues like weather and network conditions that introduce data noise. 
Furthermore, as shown in Figure~\ref{trainning}, training these RL algorithms in the real world is unstable and   cause congestion. 
To the best of our knowledge, this is the first deployment of the TSC algorithm in real cities. Consequently, city traffic management authorities mandate that the algorithms demonstrate robust noise resistance. Regarding  safety, the authorities have only allowed us to compare our algorithm with the existing expert time duration.


\noindent\textbf{(2) The RL process still needs to interact with the environment during the fuzzy duration inference process. Is this conducive to deployment?}

FuzzyLight employs a two-stage process that innovatively combines fuzzy logic with compressed sensing. The first stage ensures data integrity and reliability by using fuzzy rules to determine appropriate signal phases based on traffic flow dynamics without trial-error process. This initial process reduces the need for multiple interactions, facilitating faster setup and deployment.
%
%
Despite these advancements, the fuzzy duration inference process still requires some level of interaction with the environment. However, integrating RL with fuzzy logic has been shown to significantly accelerate the convergence of the training process (as illustrated in Figure~\ref{trainning}).

\noindent\textbf{(3) How to set the refer duration and effective range (ER) for the FuzzyLight?}

To set the reference duration in FuzzyLight, we experiment with values ranging from 30 to 50 seconds across various training datasets to assess their impact on ATT. As shown in Figure~\ref{fig:refer}, selecting a refer duration close to the mean duration of actual traffic flows is appropriate.
%
%
The impact of different ER values is also evaluated, particularly in how FuzzyLight processes different state variables like lane vehicle count and lane queue count. As shown in Figures~\ref{er1} and \ref{er2}, reducing ER leads to a decrease in the ATT. This decline is mainly due to the fact that the reduced data does not fully capture intersection conditions, highlighting the importance of a properly set ER to ensure comprehensive data utilization and optimal algorithm performance.

\noindent\textbf{(4) What are the differences between the real world and the simulated world?}

In the real world, factors such as weather and network transmission could lead to sensor delays and disconnections. Additionally, we have implemented safety modules for pedestrians and cyclists to ensure they have the shortest green light time when crossing the street. However, the behavior of violating drivers is still constrained by traffic laws. Furthermore, traffic throughput varies at different times, and traffic flow changes could be instantaneous. These sudden fluctuations could lead to intersection overflow problems, where vehicles queue at the intersection waiting to exit. These issues distinguish the real world from simulation-based environments.

\noindent\textbf{(5) How does the performance vary under different levels and types of noise?}

We have unintentionally exposed the sensors within Gaussian and U-rand noise in different noise levels, the result is shown in Table~\ref{table:guassian_levels} and Table~\ref{table:urand}.
It can be seen that the ATT performance remains good at a max scale of 100\% and below. As the noise scale increases, the ATT performance gradually declines. These experiments help illustrate how our model adapts to different noise levels, enhancing its robustness and applicability to real-world environments.
\begin{table}[h]
\centering
\caption{Gaussian noise in different noise levels.}
\label{table:guassian_levels}
\begin{tabular}{cccccc} 
\hline
\multirow{2}{*}{\begin{tabular}[c]{@{}c@{}}Scale \\(Max\_scale)\end{tabular}} & \multicolumn{3}{c}{JiNan} & \multicolumn{2}{c}{HangZhou}  \\ 
\cline{2-6}
                                                                              & 1      & 2      & 3       & 1      & 2                    \\ 
\hline
50\%                                                                          & 262.59 & 239.49 & 235.90  & 281.96 & 303.33               \\ 
\hline
100\%                                                                         & 264.39 & 251.38 & 254.86  & 284.78 & 303.56               \\ 
\hline
150\%                                                                         & 268.54 & 252.50 & 256.88  & 298.80 & 316.18               \\ 
\hline
200\%                                                                         & 278.30 & 261.67 & 255.54  & 314.41 & 323.80               \\ 
\hline
300\%                                                                         & 299.51 & 277.79 & 269.69  & 327.95 & 337.81               \\ 
\hline
500\%                                                                         & 328.94 & 303.41 & 291.83  & 359.53 & 358.79               \\
\hline
\end{tabular}
\end{table}
\begin{table}[h]
\centering
\caption{U-rand noise in different noise levels.}
\label{table:urand}
\begin{tabular}{cccccc} 
\hline
\multirow{2}{*}{\begin{tabular}[c]{@{}c@{}}Scale \\(Max\_scale)\end{tabular}} & \multicolumn{3}{c}{JiNan} & \multicolumn{2}{c}{HangZhou}  \\ 
\cline{2-6}
                                                                              & 1      & 2      & 3       & 1      & 2                    \\ 
\hline
50\%                                                                          & 264.29 & 238.79 & 237.03  & 287.09 & 307.25               \\ 
\hline
100\%                                                                         & 270.14 & 261.34 & 265.13  & 294.53 & 310.43               \\ 
\hline
300\%                                                                         & 275.93 & 273.19 & 274.63  & 305.08 & 323.53               \\ 
\hline
500\%                                                                         & 292.20 & 276.85 & 278.77  & 328.32 & 336.35               \\ 
\hline
700\%                                                                         & 311.45 & 286.73 & 284.09  & 338.19 & 348.80               \\
\hline
\end{tabular}
\end{table}

\noindent\textbf{(6) Why not using PPO and TD3 and how is their performance?}
We have supplemented the TD3 and PPO experiments in simulator, the results as shown in Table~\ref{table:other_methods}.
\begin{table}[h]
\centering
\caption{TD3 and PPO experiments.}
\label{table:other_methods}
\begin{tabular}{cccccc} 
\hline
\multirow{2}{*}{Methods} & \multicolumn{1}{l}{\multirow{2}{*}{Condition}} & \multicolumn{2}{c}{JiNan}                                                   & \multicolumn{2}{c}{HangZhou}  \\ 
\cline{3-6}
                         & \multicolumn{1}{l}{}                           & 1                                    & 3                                    & 1      & 2                    \\ 
\hline
\multirow{2}{*}{PPO~\cite{schulman2017proximal}}     & UN                                             & 261.34                               & 251.63                               & 279.53 & 301.32               \\ 
\cline{2-6}
                         & NF                                             & 251.32                               & 231.64                               & 275.14 & 295.65               \\ 
\hline
\multirow{2}{*}{TD3~\cite{fujimoto2018addressing}}     & UN                                             & 260.96                               & 250.16                               & 281.31 & 299.64               \\ 
\cline{2-6}
                         & NF                                             & \textcolor[rgb]{0.2,0.2,0.2}{250.31} & \textcolor[rgb]{0.2,0.2,0.2}{230.47} & 274.26 & 296.14               \\
\hline
\end{tabular}
\end{table}
It can be demonstrated that the performance is improved. The reason we chose to use DDPG in the real world is its simplicity, but we plan to implement these methods in the future.

\noindent More discussions are in \textbf{Appendix ~ \ref{discuss}}.
\vspace{-0.2cm}
\section{Conclusion}
\label{Sec:Conclu}
In this paper, we propose FuzzyLight, a novel two-stage approach that outputs signal phase decisions and corresponding phase durations simultaneously. FuzzyLight leverages compressed sensing technology to effectively mitigate the impact of noise interference during transmission, showcasing robustness and reliability. Meanwhile, it integrates fuzzy logic with deep reinforcement learning to determine optimal phase durations.
Our experimental results also confirm that FuzzyLight achieves state-of-the-art performance in real-world and simulated environments, maintaining high performance even under conditions of transmission noise.
In the future, we aim to optimize our method to further enhance the effectiveness and efficiency of TSC systems in practical applications.

\section*{Acknowledgements}
This work is supported by the Urban Traffic Smart Computing Body Research Project State-owned Capital Operation Budget Supporting Provincial Enterprises' Scientific Research Projects(Grant Nos. 2023GZ012) and the scientific research results transformation project.  We are grateful to Silk Road Information Port Co., Ltd., Cloud Gansu Technology Co., Ltd., and Beijing Xiaocheng Intelligent Computing Technology Co., Ltd. for providing the research platform that facilitated this work. We would also like to extend our sincere thanks to the Lanzhou Public Security Bureau and the Qingyang Public Security Bureau for their strong support and collaboration.

\clearpage
\normalem
\balance
\bibliographystyle{ACM-Reference-Format}
\bibliography{acmart}

\end{sloppypar}

\clearpage

\appendix

\section{Experiment Settings}
\label{app:settings}

%
The specific experiment settings on hyperparameters are shown in Table~\ref{table: hyperparameter}.
mart\begin{table}[H]
\centering
\caption{Hyperparameters}
\label{table: hyperparameter}
\begin{tabular}{cc} 
\toprule
Hyperparameter        & Setting        \\ 
\hline
$v_l$                 & 6              \\
$safe_l$              & 2              \\
ER                    & 160            \\
k                     & 4              \\
z                     & 20             \\
$\epsilon$            & mean=1, var=2  \\
low                   & 1              \\
high                  & 40             \\
actor\_lr             & 1e-5           \\
critic\_lr            & 2e-3           \\
refer duration        & 40             \\
discount($\gamma$)    & 0.8            \\
target critic($\tau$) & 0.95           \\
buffer capacity       & 12000          \\
epochs                & 200            \\
batch\_size           & 20             \\
learning\_rate        & 0.001          \\
target update time    & 5              \\
normal factor         & 20             \\
loss function         & MSE            \\
optimizer             & Adam           \\
\bottomrule
\end{tabular}
\end{table}

\section{Model Details}
\label{Details of the network}
We detail the design of the network structures of Fuzzy Duration Inference for FuzzyLight (as shown in Table\ref{table:laneEmbeddingDesign}).

\begin{table}[H]
\centering
\caption{Neural Network Fuzzy Membership Function}
\label{table:laneEmbeddingDesign}
\begin{tabular}{cc} 
\toprule
\multirow{2}{*}{Layer name} & Lane embedding~                                                                                               \\
                            & (input size, output size)~ ~                                                                                  \\ 
\hline
\# Lane embedding           & \begin{tabular}[c]{@{}c@{}}\\MLP(12x4x1, 12x4x4)\\Reshape(12x4x4, 12x16)\\Split(12x16, 1x16)\end{tabular}     \\ 
\hline
\# Lane attention           & \begin{tabular}[c]{@{}c@{}}\\Concat(1x16x2, 2x16)\\MHA(2x16, 2x16)\\Mean(2x16, 1x16)\end{tabular}             \\ 
\hline
\#~Lane fusion              & \begin{tabular}[c]{@{}c@{}}\\Concat(1x16x4, 4x16)\\Multiply(1x4, 4x16, 1x16)\\Reshape(1x16, 16)\end{tabular}  \\ 
\hline
\# MLP units                & \begin{tabular}[c]{@{}c@{}}\\(16, 16) \\(16, 16)\end{tabular}                                                 \\
\bottomrule
\end{tabular}
\end{table}

In addition, we detail the design of the network structures of the RL module for FuzzyLight (as shown in Table \ref{table:rlNetworkDesign}).

\begin{table}[H]
\centering
\caption{RL network design}
\label{table:rlNetworkDesign}
\begin{tabular}{ccc} 
\toprule
\multirow{2}{*}{Layer name} & Actor                                                                   & Critic                                                                                                          \\
                            & \multicolumn{2}{c}{(input size, output size)~ ~}                                                                                                                                          \\ 
\hline
\\
\# MLP units                & \begin{tabular}[c]{@{}c@{}}(16,256) \\(256, 256)\\(256, 1)\end{tabular} & \begin{tabular}[c]{@{}c@{}}(16, 16) \\(16, 32)\\Concat(32x2, 64)\\(64, 256)\\(256, 256)\\(256, 1)\end{tabular}  \\
\bottomrule
\end{tabular}
\end{table}

\section{More Experiments}
\label{app:exp}

We conduct other fuzzy logical experiments, the results are shown in Table~\ref{table:fuzzy_base} and 
we conduct experiments with eight phases in the simulated environment. 
As shown in Table~\ref{table:4}, our approach achieves SOTA performance in the JiNan2, JiNan3, and HangZhou2 datasets.

We also show the running speed of FuzzyLight in Table~\ref{tab:time}. Our proposed FuzzyLight has performed best.

\begin{figure}[h]
        \centering \includegraphics[width=0.9\linewidth]{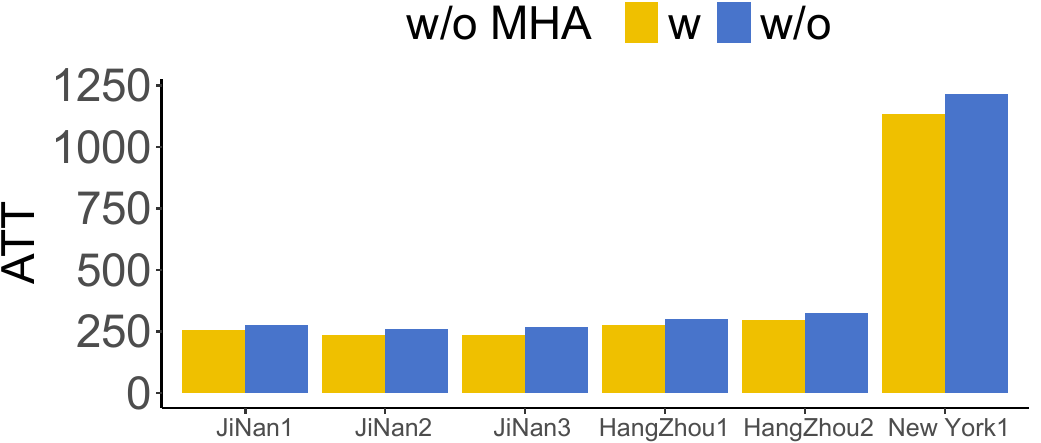}
    \caption{Ablation study for w/o using MHA layer.}
  \label{fig:attent}
\end{figure}

Moreover, to verify the effectiveness of the attention mechanism, we conduct comparison experiments. We use the network structure with the attention layer removed (w/o) for comparison, the result is shown in Figure \ref{fig:attent}.
The result shows that, after removing the attention mechanism, the ATT increases in all five datasets, with an average increase of about 3\% compared to the optimal result. It may due to different parts of the road have different effects on the phase time. 
A shorter travel time is required for the portion of the road closer to the intersection, whereas a longer travel time is required for the portion of the road further from the intersection. 
The attention module can learn the weight information between different parts of the road to better allocate the phase time dynamically.  

\begin{table}[h]
\centering
\caption{Other fuzzy logical experiments, UN denotes Under-Noise and NF denotes Noise-Free. ATT the smaller the better.}
\label{table:fuzzy_base}
\begin{tabular}{cccccc} 
\hline
Method             & Condition & JN1    & JN3    & HZ1    & HZ2     \\ 
\hline
\multirow{2}{*}{\cite{chiu1992adaptive}} & UN        & 345.13 & 341.36 & 375.32 & 391.13  \\ 
\cline{2-6}
                   & NF        & 291.34 & 283.14 & 310.34 & 378.34  \\ 
\hline
\multirow{2}{*}{\cite{chala2024intelligent}} & UN        & 324.13 & 314.36 & 356.32 & 384.13  \\ 
\cline{2-6}
                   & NF        & 286.34 & 278.14 & 301.34 & 361.34  \\
\hline
\end{tabular}
\end{table}

\begin{table}[h]
\centering
\caption{Performance of all methods in JiNan, HangZhou in terms of average travel time for eight phases. For ATT (in seconds), the smaller the better.}
\label{table:4}
\scalebox{0.9}{
\begin{tabular}{cc} 
\toprule
\textbf{Method}             & \textbf{Average waiting time}  \\ 
\hline
\textbf{\textbf{JiNan1}}    & 246.32 ± 1.33                  \\
\textbf{\textbf{JiNan2}}    & 229.93 ± 1.60                  \\
\textbf{\textbf{JiNan3}}    & 229.65 ± 1.41                  \\
\textbf{\textbf{HangZhou1}} & 274.68 ± 0.00                  \\
\textbf{\textbf{HangZhou2}} & 305.51 ± 2.77                  \\
\textbf{\textbf{New York1}} & 1211.09 ± 0.00                 \\
\bottomrule
\end{tabular}
}
\end{table}
\begin{table}[h]
\centering
\caption{Comparison on running speed (millisecond).}
\label{tab:time}
\begin{tabular}{cc} 
\hline
      Method               & Running Time  \\ 
\hline
MPLight              & $135.55$      \\
CoLight              & $155.22$      \\
Advanced-CoLight     & $99.21$       \\
CQL~\cite{CQL}                  & $77.31$       \\
TD3-BC~\cite{TD3}               & $204.05$      \\
BEAR~\cite{BEAR}                 & $211.31$      \\
Decision Transformer~\cite{decisionTransformer} & $35.62$       \\
TransformerLight~\cite{wu2023transformerlight}     & $16.56$       \\
FuzzyLight           & $10.34$       \\
\hline
\end{tabular}
\end{table}

\section{More Discussion}
\label{discuss}
\textbf{1. Why not using offline RL algorithm for TSC deployment?}

Current offline RL algorithms typically output phase but not phase duration. 
Using offline algorithms requires collecting online decision data, which can only be generated with FixedTime or other traditional methods and cannot be generated from the online RL algorithm. 
Training offline RL algorithms with such data theoretically cannot surpass FixedTime performance, resulting in suboptimal performance. 
Offline-to-online algorithms can maintain suboptimal performance during the offline phase while achieving higher performance when deployed and interacting with the environment. In the future, we aim to explore offline-to-online algorithms that can be deployed in real-world scenarios.

\noindent\textbf{2. How the FuzzyLight affects the throughput rate?}
\par First, we use the following formula to represent the throughput rate of an intersection:
 
\begin{gather}
      Throughput = \frac{x(l)}{t_{duration}}, l \in \mathcal{L}_i^{in} 
      \label{eq61}
\end{gather}
where $t_{duration}=\frac{dis}{v}$, $dis$ represents the distance of the last vehicle from intersections and $v$ denotes the average speed of the vehicle. We Then the throughput rate of a phase pair can be expressed as 
\begin{equation*}
Throughput^{i-j}=\frac{x(l_i)+x(l_j)}{max(t_{duration}^i,t_{duration}^j)}
\end{equation*}
where $l_i,l_j$ is the lane corresponding the phase pair. We assume that the last car through the intersection time is $t_{duration}=x(l)$,
\begin{equation*}
Throughput^{i-j}=\begin{cases}
1 + \frac{x(l_i)}{x_(l_j)},&x(l_i)<x(l_j);\\
1 + \frac{x(l_j)}{x_(l_i)},&x(l_i)>x(l_j).
\end{cases}
\end{equation*}
if $x(l_i)=x(l_j)$ then $Throughput=2$, therefore $1<=Throughput<=2$.

\par
\begin{figure}[H]
    \centering
    \includegraphics[width=0.5\textwidth]{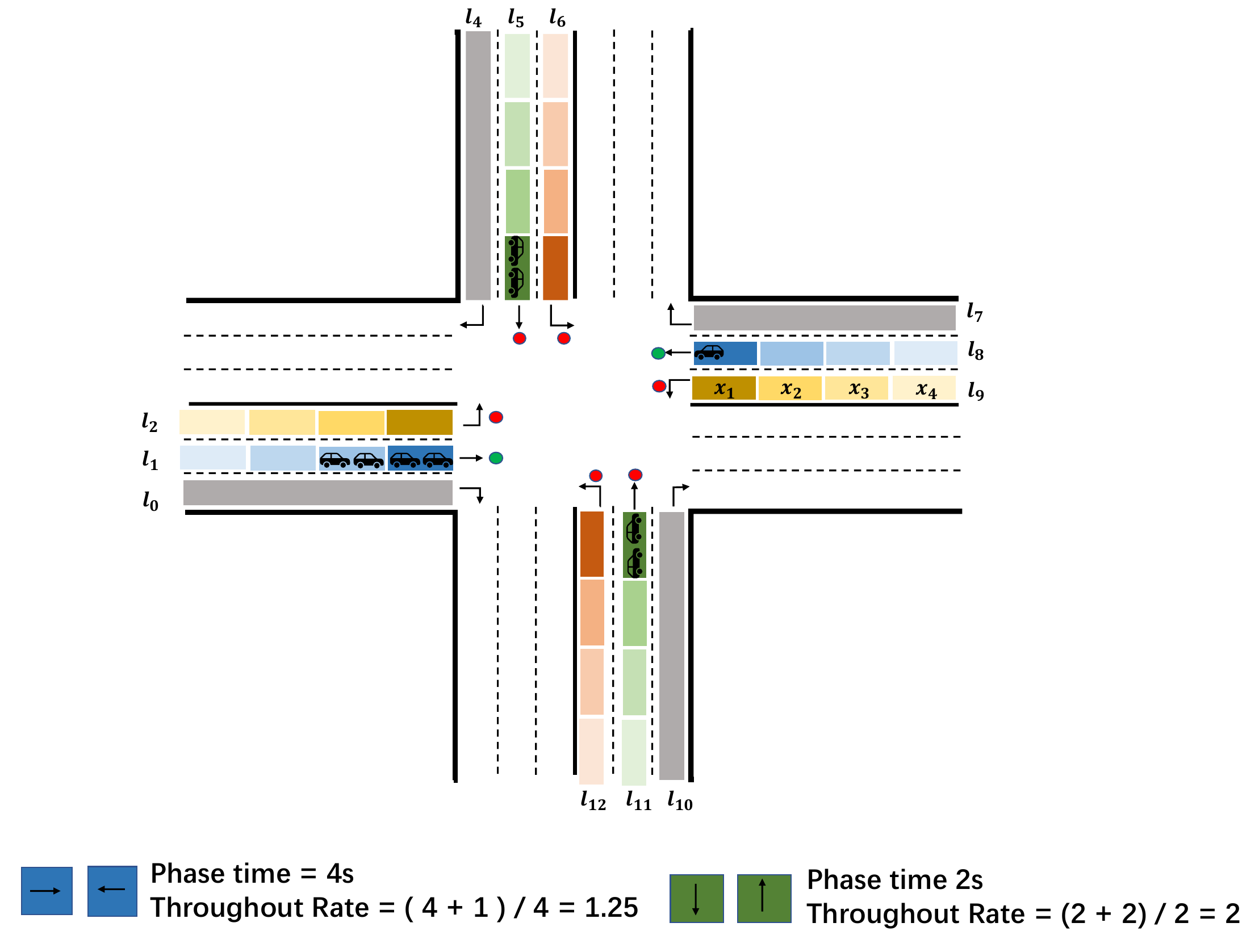}
        \caption{Vehicles in different segments of the road have different effects on phase times, with vehicles further from the intersection requiring longer green times.}
      \label{fig:segmentAttention}
\end{figure}

 As shown in Figure  \ref{fig:segmentAttention}. We have divided each road into four segments and the number of vehicles in each segment is represented as $x_i(l)$. All vehicles in the E-W and S-N phases need to be set to 4s and 2s to pass the intersection. So the $Throughput=(4+1)/4=1.25$  and $Throughput^{S-N}=(2+2)/2=2$. Our FuzzyLight model can  learn the effect of different parts of the road on the phase time, set the appropriate phase time for different phases and improve the throughput rate of the intersection. 

\par For example, in the above figure, we set the E-W phase time to 2s, and after switching to the S-N phase is also set to 2s. The throughput rate in 4s is $Q_{throughput}=(2+1)+(2+2)/(2+2)=1.75$ and compared to the previous E-W phase in 4s throughput rate of 1.25 improved by 40\%.

\end{document}